\documentclass[12pt,a4paper]{article}%
\usepackage{amsmath}
\usepackage{amsfonts}
\usepackage{amssymb}
\usepackage{graphicx}%
\begin{document}

\author{J. Jenkins$^{1}$, D. Johnson$^{2}$, L. La Ragione$^{1}$\thanks{Present
Address: Dipartimento di Ingegneria Civile e Ambientale,
Politecnico di Bari, Bari, Italy}, and H.
Makse$^{2}$\thanks{Present Address: The Levich Institute, City
College of New York, New York, NY}\\$^{1}$Department of
Theoretical and Applied Mechanics\\Cornell University, Ithaca, NY
14853\\$^{2}$Schlumberger-Doll Research\\Old Quarry Road,
Ridgefield, CT 06877 }
\title{Fluctuations and the Effective Moduli of an Isotropic, Random Aggregate of
Identical, Frictionless Spheres}
\date{Keywords: A. vibrations; B. granular materials, rock, stress waves; C.
probability and statistics. Journal: J. Mech. and Phys. of Sol.
{\bf 53}, 197-225 (2005).} \maketitle

\begin{abstract}
We consider a random aggregate of identical frictionless elastic
spheres that has first been subjected to an isotropic compression
and then sheared. We assume that the average strain provides a
good description of how stress is built up in the initial
isotropic compression. However, when calculating the increment in
the displacement between a typical pair of contaction particles
due to the shearing, we employ force equilibrium for the particles
of the pair, assuming that the average strain provides a good
approximation for their interactions with their neighbors. The
incorporation of these additional degrees of freedom in the
displacement of a typical pair relaxes the system, leading to a
decrease in the effective moduli of the aggregate. The
introduction of simple models for the statistics of the ordinary
and conditional averages contributes an additional decrease in
moduli. The resulting value of the shear modulus is in far better
agreement with that measured in numerical simulations.

\end{abstract}

\section{Introduction}

Digby (1981) and Walton (1987) considered a random aggregate of
frictional spheres in which the distribution of contacts was
isotropic. They considered a random aggregate of identical spheres
that was first compressed by an average pressure $p.$ They assumed
that the relative displacement of the centers of two contacting
particles was given by the average strain, and they obtained
expressions for the effective shear modulus $\mu^{E}$ and Lam\'{e}
coefficient
$\lambda^{E}$ . Their expressions for these moduli are \newline%
\begin{equation}
\mu^{E}=\frac{kv}{5\pi}\frac{\mu}{(1-\nu)}\left[
\frac{3\pi}{2}\frac{(1-\nu )}{vk}\frac{p}{\mu}\right]
^{1/3}\frac{[2-\nu+3\alpha(1-\nu)]}{(2-\nu)}
\label{mu}%
\end{equation}
and
\begin{equation}
\lambda^{E}=\frac{kv}{5\pi}\frac{\mu}{(1-\nu)}\left[  \frac{3\pi}{2}%
\frac{(1-\nu)}{vk}\frac{p}{\mu}\right]
^{1/3}\frac{[2-\nu-2\alpha(1-\nu
)]}{(2-\nu)}, \label{lambda}%
\end{equation}
where $k$ is the average number of contacts per particle (the
coordination number) and $v$ is the solid volume fraction. The
parameter $\alpha$ describes the strength of the transverse
stiffness of the grain-to-grain contact; $\alpha=0$ is appropriate
to frictionless interactions (perfect slip), whereas $\alpha=1$
describes the fully frictional interactions (perfect stick). The
effective bulk modulus, $\kappa^{E}$, and the effective Poisson
ratio, $\nu^{E}$, are given in terms of these by
\begin{equation}
\kappa^{E}\equiv\lambda^{E}+\frac{2}{3}\mu^{E}=\frac{kv}{3\pi}\frac{\mu
}{(1-\nu)}\left[
\frac{3\pi}{2}\frac{(1-\nu)}{vk}\frac{p}{\mu}\right]  ^{1/3}
\label{kappa}%
\end{equation}
and
\begin{equation}
\nu^{E}=\frac{\lambda^{E}}{2(\lambda^{E}+\mu^{E})}, \label{Poisson}%
\end{equation}
respectively. Note that the bulk modulus, $\kappa^{E}$, does not
depend upon $\alpha$ because the transverse forces do not enter at
all into this average strain approximation.

The effective moduli of the corresponding aggregate of
frictionless spheres can be obtained simply from equations
(\ref{mu}) and (\ref{lambda}) by setting
$\alpha=0$: \newline%
\begin{equation}
\mu^{E}=\lambda^{E}=\frac{3}{5}\kappa^{E}=\frac{kv}{5\pi}\frac{\mu}{(1-\nu
)}\left[  \frac{3\pi}{2}\frac{(1-\nu)}{vk}\frac{p}{\mu}\right]
^{1/3}.
\label{fless}%
\end{equation}
The equality of the coefficients is consistent with Cauchy's use
of the average strain assumption to obtain a single independent
modulus for random arrays of grains that interact through central
forces (e.g., Love (1927), Note B).\qquad

Jenkins, et al. (1989) compared the predicted values of the
effective shear and bulk moduli of frictional spheres with the
results of computer simulations and physical experiments on a
binary mixture of glass spheres with rather large differences in
diameters that were isotropically compressed to an average
pressure of $138$ kPa. They found that the effective shear and
bulk modulus predicted by Digby (1981) and Walton (1987) were,
respectively, three times and one and one-half times greater than
the values measured in the experiments and the simulations. Makse,
et al. (1999) also compared values of the effective moduli with
the results of computer simulations. For a binary mixture of
frictional spheres that differed little in diameter, they found
the effective shear modulus to be about two-thirds of the value
predicted using the average strain assumption. They explained this
difference as being due to the relaxation of the particles
associated with their achieving equilibrium in the numerical
simulation. More surprisingly, they found that the effective shear
modulus for frictionless spheres in the simulation was less than
ten per cent of the predicted effective medium value, equation
(\ref{fless}). It is the central goal of the present article to
understand why this is so. (By contrast the bulk modulus agreed
reasonably well with equation (\ref{kappa}) regardless whether
there was perfect slip or perfect stick.)

Because the difficulty with the shear modulus was shown to be due
to the relaxation of the particles from the average strain, we
first perform the simplest investigation that allows for some
relaxation. From the simulations, we know the rest positions of
each of the particles, as well as the vectors between the centers.
Consider a specific particle. We make the approximation that when
a macroscopic strain increment is applied, the particles in
contact with it move according to the average strain. Because the
specific particle is not, in general, in a symmetric environment,
it experiences an unbalanced force. Consequently, it will, in
general, move to a position different than that expected from the
average strain, in order to reduce the net force to zero. So, for
the specific particle, we calculate its new position. We next
calculate the energy stored within each of the contact
\textquotedblleft springs" for each of the particles in the
simulation and calculate the total stored energy due to the
applied strain. We set this equal to the usual expression for
strain energy and deduce new estimates for the bulk and shear
moduli of the aggregate. This procedure is detailed in Appendix A.

It is obvious that such a procedure can only reduce the moduli
relative to those of the average strain prediction. Note that if
we were to neglect the relaxation of each particle and assume each
that particle sees the same coordination number, $k$, distributed
uniformly around it, we would reproduce the average strain
predictions, equations (\ref{mu}) and (\ref{kappa}), as detailed
in Norris and Johnson (1997). We find that for a static confining
pressure of $100$ kPa, there is a small reduction of the bulk
modulus from $223$ MPa predicted by the average strain analysis to
$206$ MPa. There is a much larger reduction of the shear modulus
from $134$ MPa to $100$ MPa; however, the results of the
simulations for the shear modulus give the much, much smaller
value $\mu^{E}=8\pm3$ MPa. We see that relaxation effects at the
single particle level, while significant, are by no means
sufficient to explain the effect.

We are thus led to consider a more sophisticated theory in which
we explicitly account for fluctuations in pairs of contacting
particles. Here, we specialize specifically to the frictionless
case, where the reduction in shear modulus is most dramatic and
for which we can derive an analytic result using some fairly weak
assumptions. We employ the relatively simple model of a static
aggregate introduced by Jenkins (1997) in which the assumption
that the increment in the contact displacement is given by the
increment of the average strain is relaxed. Instead, the centers
of a typical pair of contacting particles are assumed to be able
to translate in order to equilibrate force, while their
surrounding neighbors are constrained to move with the increment
in the average strain. Incremental strains are employed because
the contact forces are nonlinear functions of the displacement.
When the equilibrium equations are phrased in terms of increments
in displacements, they provide linear equations for their
determination in terms of the increment in average strain. We
obtain an approximate analytical solution to the equilibrium
equations that determine the increments in displacements of the
pair in terms of the increment of average strain. This solution
contains quantities that involve the geometry and interactions of
a particle with its neighbors; we provide relatively simple
statistical models for the averages of these and their
correlations. This permits the calculation of the increment of
contact force between the pair. Summation of the increments of
contact force over all pair orientations provides the relationship
between the increment of stress and the increment in strain and,
hence, the effective moduli. In this way, we calculate a decrease
of the effective shear modulus of about seventy per cent from the
value predicted by the more elementary theory.

\section{Theory}

We focus our attention on a pair of contacting spheres, label them
$A$ and $B $, and denote the vector from the center of $A$ to the
center of $B$ by $\mathbf{d}^{(BA)}.$ We write the increment
$\mathbf{\dot{F}}^{(BA)}$in the contact force exerted by particle
$B$ on particle $A$ in terms of the increment
$\mathbf{\dot{u}}^{(BA)}$in the relative displacement of the
points of contact:
\begin{equation}
\dot{F}_{i}^{(BA)}=K_{ij}^{(BA)}\dot{u}_{j}^{(BA)}, \label{Force}%
\end{equation}
where $\mathbf{K}^{(BA)}$ is the contact stiffness.

Here, we assume that the contact stiffness is given in terms of
the unit vector $\mathbf{\hat{d}}^{(BA)}$ in the direction of
$\mathbf{d}^{(BA)}$ by
\begin{equation}
K_{ij}^{(BA)}=K_{N}^{(BA)}\hat{d}_{i}^{(BA)}\hat{d}_{j}^{(BA)},
\label{Stiff_1}%
\end{equation}
where $K_{N}^{(BA)}$ is the normal contact stiffness, given in
terms of the contact displacement $\mathbf{u}^{(BA)}$ by
\begin{equation}
K_{N}^{(BA)}=\frac{\mu d^{1/2}}{\left(  1-\nu\right)  }\left[
\delta
^{(BA)}\right]  ^{1/2}, \label{k_en}%
\end{equation}
with
\[
\delta^{(BA)}\equiv-\hat{d}_{i}^{(BA)}u_{i}^{(BA)}.
\]
Using the Hertz contact law, the normal displacement can be
related to average pressure $p$ through the average strain
assumption (Jenkins, et al., 1989) by
\begin{equation}
\delta^{(BA)}=d\left[  \frac{3\pi}{2}\frac{\left(  1-\nu\right)
}{vk}\frac
{p}{\mu}\right]  ^{2/3}, \label{delta}%
\end{equation}
where $d$ is the sphere diameter. Because computer simulations
(Jenkins, et al., 1989) indicate that the bulk modulus is rather
well predicted by the average strain theory, we believe it
appropriate to use this expression to relate the modulus $K_{N}$
to the pressure $p$ in the initial isotropic state.

The increment $\mathbf{\dot{u}}^{(BA)}$in contact displacement may
be written
in terms of the increments $\mathbf{\dot{c}}^{(B)}$ and $\mathbf{\dot{c}%
}^{(A)}$in the translations of the centers of the two spheres by
\[
\dot{u}_{i}^{(BA)}=\dot{c}_{i}^{(B)}-\dot{c}_{i}^{(A)}%
\]
Alternatively, the relative displacement of the two contacting
points may be written in terms of the increments in the averages
of quantities and their fluctuations as
\begin{equation}
\dot{u}_{i}^{(BA)}=\dot{E}_{ij}d_{j}^{(BA)}+\delta\dot{c}_{i}^{(B)}-\delta
\dot{c}_{i}^{(A)}, \label{Displacement}%
\end{equation}
where $\mathbf{\dot{E}}$ is the increment in the average strain of
the aggregate and, for example, $\delta\mathbf{\dot{c}}^{(B)}$ is
the departure of the displacement of sphere $B$ from the average
strain assumption. The relative displacement can be written more
compactly by introducing
$\mathbf{\dot{\Delta}}^{(BA)}\equiv\delta\mathbf{\dot{c}}^{(B)}-\delta
\mathbf{\dot{c}}^{(A)}$, the increment in the difference of the
fluctuations in displacement.

Given $\mathbf{\dot{F}}$, the increment $\mathbf{\dot{T}}$ in the
stress may be written as the average over all $N$ particles in a
region of relatively homogeneous strain as
\begin{equation}
\dot{T}_{ij}=\left\langle \frac{1}{V^{(A)}}\sum_{n=1}^{N^{(A)}}\dot{F}%
_{i}^{(nA)}d_{j}^{(nA)}\right\rangle
\equiv\frac{1}{2N}\sum_{A=1}^{N}\frac
{1}{V^{(A)}}\sum_{n=1}^{N^{(A)}}\dot{F}_{i}^{(nA)}d_{j}^{(nA)}, \label{Stress}%
\end{equation}
where $N^{(A)}$ is the number of particles in contact with
particle $A$ and $V^{(A)}$ is the volume of the Voroni polyhedron
associated with particle $A$. A continuous form of this may be
written in terms of a distribution function $f(\mathbf{\hat{d})}$,
defined so \ that $f(\mathbf{\hat{d})}d\Omega$ is the number of
contacts in an element $d\Omega$ of solid angle centered at
$\mathbf{\hat{d}}:$%
\[
\dot{T}_{ij}=\frac{1}{2}\frac{6v}{\pi
d^{2}}\int\int_{\Omega}f(\mathbf{\hat
{d})}\dot{F}_{i}(\mathbf{\hat{d}})\hat{d}_{j}d\Omega,
\]
where the factor multiplying the integral is the half the number
of particles per unit volume, expressed in terms of the solid
volume fraction $v $. For an isotropic distribution of contacts,
the distribution can be expressed in terms of the coordination
number:
\[
f(\mathbf{\hat{d})=}\frac{k}{4\pi}%
\]

Given the increments $\mathbf{\dot{E}}$ in average strain, the
calculation of
the increment in stress requires the increment in contact force,%
\[
\dot{F}_{i}^{(BA)}=K_{N}^{(BA)}\hat{d}_{i}^{(BA)}\hat{d}_{j}^{(BA)}\left(
\dot{E}_{jm}d_{m}^{(BA)}+\dot{\Delta}_{j}^{(BA)}\right)  ,
\]
determined in terms of the fluctuations $\mathbf{\dot{\Delta}}$
for all pairs of particles in the region. These fluctuations can
be obtained obtained as solutions of the equations of balance of
force for each of the $N $ particles.

\section{Pair Fluctuations}

Here, we analyze a far simpler situation in which two contacting
particles, $A$ and $B$, have sufficient translational freedom to
satisfy force equilibrium. In order that the equilibrium equations
for the two particles determine these translations, we assume that
the other particles in contact with the pair translate with the
average deformation.

We denote the increment in the translation of center of the
$n^{th}$ neighbor of particle $A$ by $\mathbf{\dot{c}}^{(n)}$.
Then, as before,
\[
\dot{u}_{i}^{(nA)}=\dot{c}_{i}^{(n)}-\dot{c}_{i}^{(A)}.
\]
For $n\neq B,$ only the fluctuations in the translation of
particle $A$ occur; so, for these pairs, we may write
\[
\dot{c}_{i}^{(n)}-\dot{c}_{i}^{(A)}=\dot{E}_{ij}d_{j}^{(nA)}-\delta\dot{c}%
_{i}^{(A)}=\dot{E}_{ij}d_{j}^{(nA)}+\frac{1}{2}\dot{\Delta}_{i}^{(BA)}%
-\frac{1}{2}\dot{\Sigma}_{j}^{(BA)},
\]
where $\dot{\mathbf{\Sigma}}^{(BA)}\equiv\delta\mathbf{\dot{c}}^{(B)}%
+\delta\mathbf{\dot{c}}^{(A)}$ is the increment in the sum of the
fluctuations in displacement.

The equations of force equilibrium for particle $A$ are, then,
\begin{align}
0  &  =K_{N}^{(BA)}\hat{d}_{i}^{(BA)}\hat{d}_{j}^{(BA)}\left(  \dot{E}%
_{jk}d_{k}^{(BA)}+\dot{\Delta}_{j}^{(BA)}\right) \label{Equl_eqt}\\
&  +\sum_{n\neq B}^{N^{(A)}}K_{N}^{(nA)}\hat{d}_{i}^{(nA)}\hat{d}_{j}%
^{(nA)}\left(  \dot{E}_{jk}d_{k}^{(nA)}\text{ }+\frac{1}{2}\dot{\Delta}%
_{j}^{(BA)}-\frac{1}{2}\dot{\Sigma}_{j}^{(BA)}\right)  .\nonumber
\end{align}
The corresponding equilibrium equations for particle $B$ are
obtained by
interchanging $A$ and $B,$ keeping in mind that $\mathbf{d}^{(AB)}%
=-\mathbf{d}^{(BA)}$.

The equilibrium equations for the particles $A$ and $B$ lead to a
system of equations that we use to evaluate the unknown
incremental fluctuations $\dot{\Delta}^{\left(  BA\right)  }$ and
$\dot{\Sigma}^{\left(  BA\right)  }$ for the pair $BA.$ In order
to phrase the equilibrium equations in terms of the neighbors of
the individual particles of the pair, we write, for example,
\[
\sum_{n\neq B}^{N^{(A)}}K_{N}^{(nA)}\hat{d}_{i}^{(nA)}\hat{d}_{j}^{(nA)}%
d_{l}^{(nA)}\text{
}=\sum_{n=1}^{N^{(A)}}K_{N}^{(nA)}\hat{d}_{i}^{(nA)}\hat
{d}_{j}^{(nA)}d_{l}^{(nA)}\text{ }-K_{N}^{(BA)}\hat{d}_{i}^{(BA)}\hat{d}%
_{j}^{(BA)}d_{l}^{(BA)}\text{ }.
\]
We then characterize the neighborhoods of particles $A$ and $B$
through the tensors
\[
A_{ij}^{(BA)}\equiv\sum_{n=1}^{N^{(A)}}K_{N}^{(nA)}\hat{d}_{i}^{(nA)}\hat
{d}_{j}^{(nA)}\text{ and }A_{ij}^{(AB)}\equiv\sum_{n=1}^{N^{(B)}}K_{N}%
^{(nB)}\hat{d}_{i}^{(nB)}\hat{d}_{j}^{(nB)},
\]
and
\[
J_{ijk}^{(BA)}\equiv\sum_{n=1}^{N^{(A)}}K_{N}^{(nA)}\hat{d}_{i}^{(nA)}\hat
{d}_{j}^{(nA)}\hat{d}_{k}^{(nA)}\text{ and
}J_{ijk}^{(AB)}\equiv\sum
_{n=1}^{N^{(B)}}K_{N}^{(nB)}\hat{d}_{i}^{(nB)}\hat{d}_{j}^{(nB)}\hat{d}%
_{k}^{(nB)}.
\]
The definitions of $\mathbf{A}^{\left(  BA\right)  }$,
$\mathbf{A}^{\left( AB\right)  },$ $\mathbf{J}^{\left(  BA\right)
}$ and $\mathbf{J}^{\left( AB\right)  }$ are based on the
existence of the contact between $A$ and $B$ and it is this that
gives them their directionality.

Equation (\ref{Equl_eqt}) can be written in terms of these tensors
as
\[
dJ_{ijk}^{(BA)}\dot{E}_{jk}+\frac{1}{2}A_{ij}^{(BA)}\left(  \dot{\Delta}%
_{j}^{\left(  BA\right)  }-\dot{\Sigma}_{j}^{\left(  BA\right)
}\right)
+\frac{1}{2}K_{N}^{(BA)}\hat{d}_{i}^{(BA)}\hat{d}_{j}^{(BA)}\left(
\dot{\Delta}_{j}^{\left(  BA\right)  }+\dot{\Sigma}_{j}^{\left(
BA\right) }\right)  =0.
\]
Upon interchanging $A$ and $B$, an equivalent expression is
obtained for the force equilibrium for particle $B:$
\[
dJ_{ijk}^{(AB)}\dot{E}_{jk}-\frac{1}{2}A_{ij}^{(AB)}\left(  \dot{\Delta}%
_{j}^{\left(  BA\right)  }+\dot{\Sigma}_{j}^{\left(  BA\right)
}\right)
-\frac{1}{2}K_{N}^{(BA)}\hat{d}_{i}^{(BA)}\hat{d}_{j}^{(BA)}\left(
\dot{\Delta}_{j}^{\left(  BA\right)  }-\dot{\Sigma}_{j}^{\left(
BA\right) }\right)  =0.
\]
The summations for $\mathbf{A}^{\left(  BA\right)  }$ and
$\mathbf{A}^{\left( AB\right)  }$ involve a number of contacts
equal, on average, to the coordination number; so, up to an error
of $1/k,$ the last term in these
equations may be neglected. In this case, their solution is%
\begin{equation}
\dot{\Sigma}_{i}^{\left(  BA\right)  }=d\left[  \left(
A_{ij}^{(BA)}\right) ^{-1}J_{jkm}^{(BA)}+\left(
A_{ij}^{(AB)}\right)  ^{-1}J_{jkm}^{(AB)}\right]
\dot{E}_{km}.\label{fluc_rot}%
\end{equation}
and
\begin{equation}
\dot{\Delta}_{i}^{\left(  BA\right)  }=-d\left[  \left(
A_{ij}^{(BA)}\right) ^{-1}J_{jkm}^{(BA)}-\left(
A_{ij}^{(AB)}\right)  ^{-1}J_{jkm}^{(AB)}\right]
\dot{E}_{km}\label{fluct_transl}%
\end{equation}
Given the tensors $\mathbf{A}^{(BA)}$, $\mathbf{J}^{(BA)}$, $\mathbf{A}%
^{(AB)}$, and $\mathbf{J}^{(AB)}$, equations (\ref{fluc_rot}) and
(\ref{fluct_transl}) provide the increments in the sum and
difference of the fluctuations of the pair $BA$ in terms of the
increment in average strain.

Equations (\ref{fluc_rot}) and (\ref{fluct_transl}) apply to each
pair of
contacting particles in the assembly with orientation near $\mathbf{\hat{d}%
}^{(BA)}.$ However, detailed information regarding the tensors $\mathbf{A}%
^{-1}\mathbf{J}$ for each such pair is available only from the
numerical simulations. Consequently, we introduce an average of
such tensors. The average is taken over all pairs of particles
with their orientation in an increment of solid angle centered on
the unit vector that points in the direction indicated by the
superscripts. For example, with $\Delta \Omega^{(BA)}$ the
increment of solid angle centered on the unit vector
$\mathbf{\hat{d}}^{(BA)}$:
\[
\overline{\left(  A_{ji}^{(BA)}\right)  ^{-1}J_{imn}^{(BA)}}\equiv\frac{1}%
{M}\sum_{\mathbf{d}^{(CD)}\subset\Delta\Omega^{(BA)}}\left(  A_{ji}%
^{(CD)}\right)  ^{-1}J_{imn}^{(CD)},
\]
where $M$ is the number of pairs in the increment of solid angle.
Then, the existence of the contact between particle $A$ and $B$
provides a symmetry about the plane perpendicular to
$\mathbf{\hat{d}}^{(BA)}$:
\[
\overline{\left(  A_{ji}^{(AB)}\right)
^{-1}J_{imn}^{(AB)}}=-\overline {\left(  A_{ji}^{(BA)}\right)
^{-1}J_{imn}^{(BA)}}.
\]
With this, the average value of the sum and difference of the
fluctuations
over all pairs with orientation near $\mathbf{\hat{d}}^{(BA)}$ are%
\[
\overline{\dot{\Sigma}_{j}^{\left(  BA\right)  }}=0
\]
and
\begin{equation}
\overline{\dot{\Delta}_{j}^{(BA)}}=-2d\overline{\left(
A_{ji}^{(BA)}\right)
^{-1}J_{imn}^{(BA)}}\dot{E}_{mn}.\label{Fluct}%
\end{equation}

In what follows, we also employ the corresponding averages for the
individual tensors
\[
\overline{A_{ij}^{(BA)}}\equiv\frac{1}{M}\sum_{\mathbf{d}^{(CD)}\subset
\Delta\Omega^{(BA)}}A_{ij}^{(CD)}%
\]
and%
\[
\overline{J_{ijk}^{(BA)}}\equiv\frac{1}{M}\sum_{\mathbf{d}^{(CD)}\subset
\Delta\Omega^{(BA)}}J_{ijk}^{(CD)}.
\]
Because of the role played by $\mathbf{\hat{d}}^{(BA)}$ in the
definition of the average,
$\overline{\mathbf{A}^{(AB)}}=\overline{\mathbf{A}^{(BA)}}$ and
$\overline{\mathbf{J}^{(AB)}}=-\overline{\mathbf{J}^{(BA)}};$
while the average of $\mathbf{J}^{(BA)}$ over all orientations of
the pair is zero.

In order to evaluate the average on the right hand side of
(\ref{Fluct}), we first express the tensors involved as the sum of
an average and a fluctuation. For example,
\[
A_{ij}^{(BA)}\mathbf{=}\overline{A_{ij}^{(BA)}}\mathbf{+}A_{ij}^{(BA)\prime}.
\]
Then, in (\ref{Fluct}), we approximate the inverse of
$\mathbf{A}^{(BA)}$ in terms of the inverse of
$\overline{\mathbf{A}^{(BA)}}$and the fluctuation
$\mathbf{A}^{(BA)\prime}$ by
\[
\left(  A_{ij}^{(BA)}\right)  ^{-1}=\left[  \delta_{il}+\left(
\overline {A_{ik}^{(BA)}}\right)  ^{-1}A_{kl}^{(BA)\prime}\right]
^{-1}\left( \overline{A_{lj}^{(BA)}}\right)  ^{-1},
\]
or, up to an error proportional to the cube of the fluctuations
\begin{align*}
\left(  A_{ij}^{(BA)}\right)  ^{-1}  &  \doteq\left[
\delta_{il}-\left(
\overline{A_{ik}^{(BA)}}\right)  ^{-1}A_{kl}^{(BA)\prime}\right. \\
&  \left.  +\left(  \overline{A_{ik}^{(BA)}}\right)
^{-1}A_{km}^{(BA)\prime }\left(  \overline{A_{mp}^{(BA)}}\right)
^{-1}A_{pl}^{(BA)\prime}\right] \left(
\overline{A_{lj}^{(BA)}}\right)  ^{-1}.
\end{align*}
With this,
\begin{align}
&  \overline{A_{ji}^{(BA)-1}J_{imn}^{(BA)}}\nonumber\\
&  =\left(  \overline{A_{ji}^{(BA)}}\right)  ^{-1}\overline{J_{imn}^{(BA)}%
}-\left(  \overline{A_{jk}^{(BA)}}\right)  ^{-1}\left(  \overline
{A_{li}^{(BA)}}\right)
^{-1}\overline{A_{kl}^{(BA)\prime}J_{imn}^{(BA)\prime
}}\nonumber\\
&  +\left(  \overline{A_{jk}^{(BA)}}\right)  ^{-1}\left(
\overline {A_{sp}^{(BA)}}\right)  ^{-1}\left(
\overline{A_{li}^{(BA)}}\right)
^{-1}\overline{A_{ks}^{(BA)\prime}A_{pl}^{(BA)\prime}}\ \overline
{J_{imn}^{(BA)}} \label{final_fluc_sol}%
\end{align}
We next introduce simple assumptions regarding the distribution of
contacts and calculate these averages.

\section{Averages}

We introduce an orthogonal Cartesian system with its center
coincident with the center of the sphere $A$ and characterize a
typical contact vector $\mathbf{\alpha}$ through its equatorial
and polar angles $\phi$ and $\theta$:
$\mathbf{\alpha}=(\sin\theta\cos\phi,\sin\theta\sin\phi,\cos\theta).$

In calculating the tensors $\overline{\mathbf{A}^{(BA)}}$ and
$\overline {\mathbf{J}^{(BA)}}$, we replace the summation over the
discrete contacts of a
particle by integration over a contact distribution function $g^{(BA)}%
(\mathbf{\alpha)}$, defined so \ that
$g^{(BA)}(\mathbf{\alpha)}d\Omega$ is the number of contacts in an
element of solid angle $d\Omega=\sin\theta d\theta d\phi$ centered
at $\mathbf{\alpha}$, given that there is a contact at
$\mathbf{\hat{d}}^{(BA)}$ along the polar axis. Then, for example,
\[
\overline{A_{ij}^{(BA)}}=K_{N}\hat{d}_{i}^{(BA)}\hat{d}_{j}^{(BA)}+K_{N}%
\int_{\Omega_{\alpha}}g(\mathbf{\alpha})\alpha_{i}\alpha_{j}d\Omega_{\alpha},
\]
where the integration is over all solid angle $\Omega_{\alpha}$
consistent with the presence of particle $B.$ The first term is
associated with the presence of this contact. We assume that the
distribution is uniform in the upper and lower hemispheres. That
is, in the upper hemisphere, there are, on average, $k/2-1$
contacting particles uniformly distributed over the orientations
not excluded by the solid angle of $\pi$ associated with particle
$B$ at the pole; while there are, on average, $k/2$ contacting
particles uniformly distributed over the lower hemisphere. Then
\[
g(\mathbf{\alpha})=\left\{
\begin{array}
[c]{c}%
0,\text{ }0\leq\theta_{\alpha}\leq\frac{\pi}{3}\\
\frac{\left(  k-2\right)  }{2\pi},\text{ }\frac{\pi}{3}\leq\theta_{\alpha}%
\leq\frac{\pi}{2}\\
\frac{k}{4\pi},\text{ }\frac{\pi}{2}\leq\theta_{\alpha}\leq\pi
\end{array}
\right.  ,
\]
and
\begin{align*}
&  \overline{A_{ij}^{(BA)}}=K_{N}\hat{d}_{i}^{(BA)}\hat{d}_{j}^{(BA)}%
+K_{N}\left[
\int_{0}^{2\pi}\int_{\frac{\pi}{3}}^{\frac{\pi}{2}}\frac
{k-2}{2\pi}\alpha_{i}\alpha_{j}\sin\theta d\theta d\phi\right. \\
&  \left.
+\int_{0}^{2\pi}\int_{\frac{\pi}{2}}^{\pi}\frac{k}{4\pi}\alpha
_{i}\alpha_{j}\sin\theta d\theta d\phi\right]  .
\end{align*}
It is a straightforward calculation to determine that
\begin{equation}
\overline{A_{ij}^{(BA)}\text{ }}\equiv K_{N}\left(
\alpha_{1}\delta
_{ij}+\alpha_{2}\hat{d}_{i}^{(BA)}\hat{d}_{j}^{(BA)}\right)  , \label{ABAR}%
\end{equation}
where $\alpha_{1}=$ $(19k-22)/48$ and $\alpha_{2}=$ $(22-3k)/16.$

In a similar way, we determine that
\begin{equation}
\overline{J_{ijk}^{(BA)}}\equiv K_{N}\left[  \omega_{1}\hat{d}_{i}^{(BA)}%
\hat{d}_{j}^{(BA)}\hat{d}_{k}^{(BA)}+\omega_{2}\left(  \delta_{ik}\hat{d}%
_{j}^{(BA)}+\delta_{ij}\hat{d}_{k}^{(BA)}+\delta_{kj}\hat{d}_{i}%
^{(BA)}\right)  \right]  , \label{JBAR}%
\end{equation}
where $\omega_{1}=(166-11k)/128$ and $\omega_{2}=-(k+14)/128.$

For simplicity, we approximate the tensor
$\overline{\mathbf{A}^{(BA)}}$ by its isotropic part:
\[
\overline{A_{ij}^{(BA)}}=\psi K_{N}\delta_{ij},
\]
where $\psi=k/3$. Then its inverse is simply
\begin{equation}
\overline{A_{ij}^{(BA)}}^{-1}=(\psi K_{N})^{-1}\delta_{ij}. \label{AINV}%
\end{equation}
With this, the terms in equation (\ref{final_fluc_sol}) that
remain to be evaluated are
$\overline{\mathbf{A}^{(BA)\prime}\mathbf{J}^{(BA)\prime}}$ and
$\overline{\mathbf{A}^{(BA)\prime}\mathbf{A}^{(BA)\prime}}.$

Now, by definition,
\[
\overline{A_{ji}^{(BA)\prime}J_{imn}^{(BA)\prime}}=\overline{A_{ji}%
^{(BA)}J_{imn}^{(BA)}}-\overline{A_{ji}^{(BA)}}\
\overline{J_{imn}^{(BA)}}.
\]
In order to calculate
$\overline{\mathbf{A}^{(BA)}\mathbf{J}^{(BA)}}$ we introduce the
joint probability density function $F(\mathbf{\alpha
},\mathbf{\beta})$, defined so that $F(\mathbf{\alpha},\mathbf{\beta}%
)d\Omega_{\alpha}d\Omega_{\beta}$ is the fraction of contacts with
$\mathbf{\alpha}$ in $d\Omega_{\alpha}$ and $\mathbf{\beta}$ in
$d\Omega _{\beta} $, given that there is the contact at
$\mathbf{\hat{d}}^{(BA)}$, including the possibility that the two
directions can coincide. Then
\begin{align*}
\overline{A_{ji}^{(BA)}J_{imn}^{(BA)}}  &
=K_{N}^{2}\hat{d}_{j}^{(BA)}\hat
{d}_{m}^{(BA)}\hat{d}_{n}^{(BA)}+K_{N}^{2}\hat{d}_{i}^{(BA)}\hat{d}_{m}%
^{(BA)}\hat{d}_{n}^{(BA)}\int_{\Omega_{\mathbf{\beta}}}g(\mathbf{\beta}%
)\beta_{j}\beta_{i}d\Omega_{\beta}\\
&  +K_{N}^{2}\hat{d}_{j}^{(BA)}\hat{d}_{i}^{(BA)}\int_{\Omega_{\alpha}%
}g(\mathbf{\alpha})\alpha_{i}\alpha_{m}\alpha_{n}d\Omega_{\alpha}\\
&  +K_{N}^{2}\int_{\Omega_{\beta}}\int_{\Omega_{\mathbf{\alpha}}%
}F(\mathbf{\alpha},\mathbf{\beta})\beta_{j}\beta_{i}\alpha_{i}\alpha_{m}%
\alpha_{n}d\Omega_{\alpha}d\Omega_{\beta},
\end{align*}
where the integrals are taken over all solid angles consistent
with the presence of particle $B.$ Then
\begin{align}
\overline{A_{ji}^{(BA)\prime}J_{imn}^{(BA)\prime}}  &
=K_{N}^{2}\int
_{\Omega_{\beta}}\int_{\Omega_{\mathbf{\alpha}}}F(\mathbf{\alpha
},\mathbf{\beta})\beta_{j}\beta_{i}\alpha_{i}\alpha_{m}\alpha_{n}%
d\Omega_{\alpha}d\Omega_{\beta}\label{FlucProd}\\
&  -K_{N}^{2}\int_{\Omega_{\beta}}g(\mathbf{\beta})\beta_{j}\beta_{i}%
d\Omega\int_{\Omega_{\alpha}}g(\mathbf{\alpha})\alpha_{i}\alpha_{m}\alpha
_{n}d\Omega,\nonumber
\end{align}
in which the contributions associated with the presence of the
contact at $\mathbf{\hat{d}}^{(BA)}$ have canceled.

The unconditional averages are
\[
\int_{\Omega_{\beta}}g(\mathbf{\beta})\beta_{j}\beta_{i}d\Omega=\alpha
_{1}\delta_{ji}+\widetilde{\alpha}_{2}\hat{d}_{j}^{(BA)}\hat{d}_{i}^{(BA)},
\]
where $\alpha_{1}$ has been defined in (\ref{ABAR}) and
$\widetilde{\alpha }_{2}=\left(  18-9k\right)  /48,$ and
\begin{align}
&
\int_{\Omega_{\beta}}g(\mathbf{\alpha})\alpha_{i}\alpha_{m}\alpha
_{n}d\Omega_{\beta}\nonumber\\
&  =\widetilde{\omega}_{1}\hat{d}_{i}^{(BA)}\hat{d}_{m}^{(BA)}\hat{d}%
_{n}^{(BA)}-\omega_{2}\left(
\delta_{im}\hat{d}_{n}^{(BA)}+\delta_{mn}\hat
{d}_{i}^{(BA)}+\delta_{ni}\hat{d}_{m}^{(BA)}\right)  , \label{INTG}%
\end{align}
where $\widetilde{\omega}_{1}=\left(  38-11k\right)  /128$ and
$\omega_{2}$ is defined in (\ref{JBAR}).

The joint probability density can be expressed as the product of
the simple probability $g(\mathbf{\beta})$ and the conditional
joint probability
$h_{\alpha|\beta}(\mathbf{\alpha},\mathbf{\beta})$, defined so
that
$h_{\alpha|\beta}(\mathbf{\alpha},\mathbf{\beta})d\Omega_{\alpha}$
is the fraction of contacts with $\mathbf{\alpha}$ in
$d\Omega_{\alpha}$, given that $\mathbf{\beta}$ is in
$d\Omega_{\beta}$:
\[
F(\mathbf{\alpha},\mathbf{\beta})=g(\mathbf{\beta})h_{\alpha|\beta
}(\mathbf{\alpha},\mathbf{\beta}).
\]
Here $h_{\alpha|\beta}(\mathbf{\alpha},\mathbf{\beta})$ includes
the possibility that $\mathbf{\alpha}$ equals $\mathbf{\beta}$. We
take this possibility into account explicitly and introduce the
conditional probability
$z_{\alpha|\beta}(\mathbf{\alpha},\mathbf{\beta})d\Omega_{\alpha}$
that expresses the fraction of contacts at $\mathbf{\alpha}$ in
$d\Omega_{\alpha}$, given that $\mathbf{\beta}$ is in
$d\Omega_{\beta}$ with $\mathbf{\alpha \neq\beta.}$ Then
\begin{align}
&
\int_{\Omega_{\beta}}\int_{\Omega_{\mathbf{\alpha}}}F(\mathbf{\alpha
},\mathbf{\beta})\beta_{j}\beta_{i}\alpha_{i}\alpha_{m}\alpha_{n}%
d\Omega_{\alpha}d\Omega_{\beta}\nonumber\\
&  =\int_{\Omega_{\beta}}g(\mathbf{\beta})\beta_{j}\beta_{m}\beta_{n}%
d\Omega_{\beta}+\int_{\Omega_{\beta}}\int_{\Omega_{\alpha}}g(\mathbf{\beta
})z_{\alpha|\beta}(\mathbf{\alpha},\mathbf{\beta})\beta_{j}\beta_{i}\alpha
_{i}\alpha_{m}\alpha_{n}d\Omega_{\alpha}d\Omega_{\beta}\nonumber\\
&  \label{FAJ}%
\end{align}
The conditional probability
$z_{\alpha|\beta}(\mathbf{\alpha},\mathbf{\beta})$ is not uniform;
it depends on the locations of spheres at $\mathbf{\beta}$ and
$\mathbf{\hat{d}}^{(BA)}.$

>From the above results, we have
\begin{align}
\overline{A_{js}^{(BA)\prime}A_{sl}^{(BA)\prime}}  &
=K_{N}^{2}\int
_{\Omega_{\beta}}\int_{\Omega_{\mathbf{\alpha}}}F(\mathbf{\alpha
},\mathbf{\beta})\beta_{j}\beta_{s}\alpha_{s}\alpha_{l}d\Omega_{\beta}%
d\Omega_{\alpha}\nonumber\\
&  -K_{N}^{2}\int_{\Omega_{\beta}}g(\mathbf{\beta})\beta_{j}\beta_{s}%
d\Omega\int_{\Omega_{\alpha}}g(\mathbf{\alpha})\alpha_{s}\alpha_{l}d\Omega,
\label{Fluct_ayes}%
\end{align}
with
\begin{align}
&
\int_{\Omega_{\beta}}\int_{\Omega_{\mathbf{\alpha}}}F(\mathbf{\alpha
},\mathbf{\beta})\beta_{j}\beta_{s}\alpha_{s}\alpha_{l}d\Omega
d\Omega
\nonumber\\
&
=\int_{\Omega_{\beta}}g(\mathbf{\beta})\beta_{j}\beta_{l}d\Omega
+\int_{\Omega_{\beta}}\int_{\Omega_{\alpha}}g(\mathbf{\beta})z_{\alpha|\beta
}(\mathbf{\alpha},\mathbf{\beta})\beta_{j}\beta_{s}\alpha_{s}\alpha_{l}%
d\Omega_{\beta}d\Omega_{\alpha}. \label{Eff_A_A}%
\end{align}

The integrals in (\ref{FAJ}) and (\ref{Eff_A_A}) that involve the
conditional probability are complicated because the limits of
integration for $\theta_{\alpha}$ and $\phi_{\alpha}$ can depend
upon $\theta_{\beta}$ and $\phi_{\beta}$. We illustrate this in
the calculation of $\overline
{\mathbf{A}^{(BA)\prime}\mathbf{J}^{(BA)\prime}}$with the order of
integration taken to be $d\phi_{\alpha}$, $d\phi_{\beta}$,
$d\theta_{\alpha}$, $d\theta_{\beta}.$ Then
\begin{align}
&  \int_{\Omega_{\mathbf{\beta}}}\int_{\Omega_{\mathbf{\alpha}}}%
g(\mathbf{\beta})z_{\alpha|\beta}(\mathbf{\alpha},\mathbf{\beta})\beta
_{j}\beta_{i}\alpha_{i}\alpha_{m}\alpha_{n}d\Omega_{\alpha}d\Omega_{\beta
}\label{cond_equat}\\
&  =\int_{\frac{\pi}{3}}^{\frac{2\pi}{3}}\int_{\pi/3}^{\theta_{\beta}%
+\frac{\pi}{3}}\int_{0}^{2\pi}\int_{\Phi+\phi_{\beta}}^{2\pi-\Phi+\phi_{\beta
}}g(\mathbf{\beta})z_{\alpha|\beta}(\mathbf{\alpha},\mathbf{\beta})\beta
_{j}\beta_{i}\alpha_{i}\alpha_{m}\alpha_{n}d\Omega_{\alpha}d\Omega_{\beta
}\nonumber\\
&  +\int_{\frac{\pi}{3}}^{\frac{2\pi}{3}}\int_{\theta_{\beta}+\frac{\pi}{3}%
}^{\pi}\int_{0}^{2\pi}\int_{0}^{2\pi}g(\mathbf{\beta})z_{\alpha|\beta
}(\mathbf{\alpha},\mathbf{\beta})\beta_{j}\beta_{i}\alpha_{i}\alpha_{m}%
\alpha_{n}d\Omega_{\alpha}d\Omega_{\beta}\nonumber\\
&
+\int_{\frac{2\pi}{3}}^{\pi}\int_{\frac{\pi}{3}}^{\theta_{\beta}-\frac{\pi
}{3}}\int_{0}^{2\pi}\int_{0}^{2\pi}g(\mathbf{\beta})z_{\alpha|\beta
}(\mathbf{\alpha},\mathbf{\beta})\beta_{j}\beta_{i}\alpha_{i}\alpha_{m}%
\alpha_{n}d\Omega_{\alpha}d\Omega_{\beta}\nonumber\\
&
+\int_{\frac{2\pi}{3}}^{\pi}\int_{\theta_{\beta}-\frac{\pi}{3}}^{\frac
{5\pi}{3}-\theta_{\beta}}\int_{0}^{2\pi}\int_{\Phi+\phi_{\beta}}^{2\pi
-\Phi+\phi_{\beta}}g(\mathbf{\beta})z_{\alpha|\beta}(\mathbf{\alpha
},\mathbf{\beta})\beta_{j}\beta_{i}\alpha_{i}\alpha_{m}\alpha_{n}%
d\Omega_{\alpha}d\Omega_{\beta}.\nonumber
\end{align}
The angle $\Phi(\theta_{\alpha},\theta_{\beta})$ follows from the
fact that the arc on the unit sphere that links the centers of the
spheres at $\mathbf{\alpha}$ and $\mathbf{\beta}$ has length
$d(\mathbf{\alpha
},\mathbf{\beta})=\cos^{-1}(\mathbf{\alpha}\cdot\mathbf{\beta}).$
When $\mathbf{\alpha}$ is taken to be in the $x-z$ plane,
$\phi_{\alpha}=0$ and
$\phi_{\beta}=\Phi;$ then, because the spheres are in contact, $d=\pi/3$ and%
\[
\Phi=\arccos\left(  \frac{1-2\cos\theta_{\alpha}\cos\theta_{\beta}}%
{2\sin\theta_{\alpha}\sin\theta_{\beta}}\right)  .
\]
The distribution function
$z_{\alpha|\beta}(\mathbf{\alpha},\mathbf{\beta})$ is determined
over the range of polar angles in each of these integrals in
Appendix B. Then, in Appendix C, it is used to complete the
calculation of
$\overline{\mathbf{A}^{(BA)\prime}\mathbf{J}^{(BA)\prime}}$ and
$\overline {\mathbf{A}^{(BA)\prime}\mathbf{A}^{(BA)\prime}}$.

The final results, obtained in Appendix C, are
\begin{align}
&  \overline{A_{ji}^{(BA)\prime}J_{imn}^{(BA)\prime}}\nonumber\\
&  =K_{N}^{2}\left[  S_{jmn}^{(BA)}-\omega_{2}\alpha_{1}\left(
\delta
_{jm}\hat{d}_{n}^{(BA)}+\delta_{nj}\hat{d}_{m}^{(BA)}\right)
-\left(
\omega_{2}\widetilde{\alpha}_{2}+\omega_{2}\alpha_{1}\right)  \delta_{nm}%
\hat{d}_{j}^{(BA)}\right. \nonumber\\
&  \left.  -\left(  \alpha_{1}\widetilde{\omega}_{1}+\widetilde{\omega}%
_{1}\widetilde{\alpha}_{2}+2\omega_{2}\widetilde{\alpha}_{2}\right)
\hat
{d}_{m}^{(BA)}\hat{d}_{j}^{(BA)}\hat{d}_{n}^{(BA)}\right]  \label{AJFluct}%
\end{align}
and
\begin{equation}
\overline{A_{js}^{(BA)\prime}A_{sl}^{(BA)\prime}}=K_{N}^{2}\left[
H_{jl}^{(BA)}-\alpha_{1}^{2}\delta_{jl}-\left(
2\alpha_{1}\widetilde{\alpha
}_{2}+\widetilde{\alpha}_{2}^{2}\right)  \hat{d}_{j}^{(BA)}\hat{d}_{l}%
^{(BA)}\right]  , \label{AAFluct}%
\end{equation}
where $\mathbf{S}^{(BA)}$ and $\mathbf{H}^{(BA)}$ are given as
functions of $k$ and $\mathbf{d}^{(BA)}$ in equations (\ref{Ess})
and (\ref{aitch}), respectively, of Appendix C.

Equations (\ref{AJFluct}) and (\ref{AAFluct}) may be written more
compactly as
\begin{align}
\overline{A_{ji}^{(BA)\prime}J_{imn}^{(BA)\prime}}  &
=K_{N}^{2}\left[
\kappa_{1}\hat{d}_{m}^{(BA)}\hat{d}_{j}^{(BA)}\hat{d}_{n}^{(BA)}+\kappa
_{2}\left(  \delta_{jm}\hat{d}_{n}^{(BA)}+\delta_{nj}\hat{d}_{m}%
^{(BA)}\right)  \right. \nonumber\\
&  \left.  +\kappa_{3}\delta_{nm}\hat{d}_{j}^{(BA)}\right]  \label{kappa2}%
\end{align}
and
\begin{equation}
\overline{A_{js}^{(BA)\prime}A_{sl}^{(BA)\prime}}=K_{N}^{2}\left(
\eta
_{1}\delta_{jl}+\eta_{2}\hat{d}_{j}^{(BA)}\hat{d}_{l}^{(BA)}\right)
,
\label{Eta}%
\end{equation}
where, with the information provided, the coefficients $\kappa$
and $\eta$ may be expressed as functions of $k$; they are
evaluated for a specific value of $k$ in Appendix C. Then, the
last term of equation (\ref{final_fluc_sol}) is
\begin{align}
&  \left(  \overline{A_{jk}^{(BA)}}\right)  ^{-1}\left(  \overline
{A_{sp}^{(BA)}}\right)  ^{-1}\left(
\overline{A_{li}^{(BA)}}\right)
^{-1}\overline{A_{ks}^{(BA)\prime}A_{pl}^{(BA)\prime}}\ \overline
{J_{imn}^{(BA)}}\nonumber\\
&  =\psi^{-3}\left[  \xi_{1}\hat{d}_{j}^{(BA)}\hat{d}_{n}^{(BA)}\hat{d}%
_{m}^{(BA)}+\xi_{2}\left(
\delta_{jm}\hat{d}_{n}^{(BA)}+\delta_{jn}\hat
{d}_{m}^{(BA)}\right)
+\xi_{3}\delta_{nm}\hat{d}_{j}^{(BA)}\right]  ,
\label{FINAL}%
\end{align}
where $\xi_{1}\equiv\left(
\eta_{1}\omega_{1}+\eta_{2}\omega_{1}+2\eta _{2}\omega_{2}\right)
,$ $\xi_{2}\equiv\eta_{1}\omega_{2},$ and $\xi _{3}\equiv\left(
\eta_{2}\omega_{2}+\eta_{1}\omega_{2}\right)  $, with the
coefficients $\eta$ and $\omega$ defined in (\ref{Eta}) and
(\ref{JBAR}), respectively.

\section{Effective Moduli}

We use (\ref{Force}), (\ref{Stiff_1}), and (\ref{Displacement}) to
write the increment in contact force as
\[
\overline{\dot{F}_{i}^{(BA)}}=K_{N}^{(BA)}\hat{d}_{i}^{(BA)}\hat{d}_{j}%
^{(BA)}\left(  \dot{E}_{jm}d_{m}^{(BA)}+\overline{\dot{\Delta}_{j}^{(BA)}%
}\right)  ,
\]
where $K_{N}$ is given by (\ref{k_en}) and (\ref{delta}) and
\begin{align*}
&  \overline{\dot{\Delta}_{j}^{(BA)}}=-2\overline{A_{ji}^{(BA)-1}%
J_{imn}^{(BA)}}\dot{E}_{mn}\\
&  =-2\left\{  \psi^{-1}\left[  \omega_{1}\hat{d}_{j}^{(BA)}\hat{d}_{n}%
^{(BA)}\hat{d}_{m}^{(BA)}+\omega_{2}\left(  \delta_{jm}\hat{d}_{n}%
^{(BA)}+\delta_{jn}\hat{d}_{m}^{(BA)}\right)
+\omega_{2}\delta_{nm}\hat
{d}_{j}^{(BA)}\right]  \right. \\
&  -\psi^{-2}\left[  \kappa_{1}\hat{d}_{j}^{(BA)}\hat{d}_{n}^{(BA)}\hat{d}%
_{m}^{(BA)}+\kappa_{2}\left(
\delta_{jm}\hat{d}_{n}^{(BA)}+\delta_{jn}\hat
{d}_{m}^{(BA)}\right)  +\kappa_{3}\delta_{nm}\hat{d}_{j}^{(BA)}\right] \\
&  \left.  +\psi^{-3}\left[  \xi_{1}\hat{d}_{j}^{(BA)}\hat{d}_{n}^{(BA)}%
\hat{d}_{m}^{(BA)}+\xi_{2}\left(
\delta_{jm}\hat{d}_{n}^{(BA)}+\delta
_{jn}\hat{d}_{m}^{(BA)}\right)
+\xi_{3}\delta_{nm}\hat{d}_{j}^{(BA)}\right] \right\}
\dot{E}_{mn}.
\end{align*}

With the increment contact force known, the increment in stress,
\[
\dot{T}_{ij}=\frac{3v}{\pi
d^{2}}\frac{k}{4\pi}\int_{\Omega}\overline{\dot
{F}_{i}^{(BA)}}\hat{d}_{j}^{(BA)}d\Omega,
\]
may be calculated, making use of the identities
\[
\int_{\Omega}\hat{d}_{i}^{(BA)}\hat{d}_{j}^{(BA)}d\Omega=\frac{4\pi}{3}%
\delta_{ij}%
\]
and
\[
\int_{\Omega
i}\hat{d}_{i}^{(BA)}\hat{d}_{j}^{(BA)}\hat{d}_{k}^{(BA)}\hat
{d}_{l}^{(BA)}d\Omega=\frac{4\pi}{15}(\delta_{ij}\delta_{kl}+\delta_{jl}%
\delta_{ik}+\delta_{il}\delta_{jk}).
\]
Therefore, the incremental response of an isotropic, random
aggregate of identical frictionless spheres is given by
\begin{align*}
&  \dot{T}_{ij}\\
&  =\frac{vkK_{N}}{5\pi
d}\{2[1-2\psi^{-1}(\omega_{1}+2\omega_{2})+2\psi
^{-2}(\kappa_{1}+2\kappa_{2})-2\psi^{-3}(\xi_{1}+2\xi_{2})]\dot{E}_{ij}\\
&  +\left[  1-2\psi^{-1}(\omega_{1}+2\omega_{2})-10\psi^{-1}\omega_{2}%
+2\psi^{-2}(\kappa_{1}+2\kappa_{2})+10\psi^{-2}\kappa_{3}\right. \\
&  \left.  \left.
-2\psi^{-3}(\xi_{1}+2\xi_{2})-10\psi^{-3}\xi_{3}\right]
\dot{E}_{kk}\delta_{ij}\right\}  .
\end{align*}
>From this, the effective moduli are
\begin{equation}
\mu^{E}=\frac{kv}{5\pi d}K_{N}\{1-2\left[
\psi^{-1}(\omega_{1}+2\omega
_{2})-\psi^{-2}(\kappa_{1}+2\kappa_{2})+\psi^{-3}(\xi_{1}+2\xi_{2})\right]
\}
\label{shear_hetero}%
\end{equation}
and
\begin{equation}
\lambda^{E}=\frac{kv}{5\pi d}K_{N}[1-2\psi^{-1}(\omega_{1}+7\omega_{2}%
)+2\psi^{-2}(\kappa_{1}+2\kappa_{2}+5\kappa_{3})-2\psi^{-3}(\xi_{1}+2\xi
_{2}+5\xi_{3})], \label{Lambda_hetero}%
\end{equation}
where, we recall that $\psi=k/3$ and the coefficients $\kappa$,
$\omega$, and $\xi$ are defined in (\ref{kappa2}), (\ref{JBAR}),
and (\ref{FINAL}), respectively. The dependence of the effective
moduli on the pressure occurs only through $K_{N}$ and $k$, the
other coefficients are functions only of the geometry. A
dependence on $p^{1/3}$ enters through $K_{N}.$

\section{Comparisons}

We compare the results of computer simulations with the
predictions of the model. The numerical simulations were carried
out using the code TRUBAL developed by Cundall (1988). Each
simulation employed spheres of two different radii:
$R_{1}=0.105\times10^{-3}$ \ m and $R_{2}=0.095\times10^{-3}$ \ m
in equal numbers. The shear modulus of the material of the spheres
was $\mu=2.9\times10^{10}$ Pa and the Poisson ratio was $\nu=0.2$.
This simulation employed a system of $10,000$ spheres. The initial
state was obtained in the manner described by Makse, et al.
(1999). An initial random aggregate of frictionless spheres
without contacts was homogeneously and isotropically contracted,
bringing the spheres into contact, until a pressure $p$ of
$100\times10^{3}$ Pa was reached. In this state, $k=6.067$ and
$v=0.637.$

The numerical results for the shear and bulk moduli were
\[
\mu^{E}=8\text{ MPa and }\kappa^{E}=200\text{ MPa}.
\]

The predictions are based on identical spheres made of the same
material with the average diameter $d=0.1995\times10^{-3}$ \ m in
an initial state with the same coordination number, volume
fraction, and confining pressure. Then, from equations
(\ref{k_en}) and (\ref{delta}), we first calculate the average
normal stiffness:
\[
K_{N}=1.08\times10^{5}\text{ Pa-m.}%
\]
In this case, the effective moduli calculated from the average
strain
assumption are%
\[
\mu^{E}=\lambda^{E}=134\text{ MPa},
\]
and%
\[
\kappa^{E}=\lambda^{E}+\frac{2}{3}\mu^{E}=\frac{5}{3}\mu^{E}=223\text{
MPa}.
\]
On the other hand, the effective moduli determined from equations
(\ref{shear_hetero}) and (\ref{Lambda_hetero}) are
\[
\mu^{E}=\frac{kv}{5\pi d}K_{N}(1-0.69)\text{ }=41\text{ MPa}%
\]
and
\[
\lambda^{E}=\frac{kv}{5\pi d}K_{N}(1+0.57)=208\text{ MPa};
\]
so, the bulk modulus is
\[
\kappa^{E}=\text{ }235\text{ MPa.}%
\]
The predicted value of the effective shear modulus is about $70\%$
less than that resulting from the average strain assumption. This
is encouraging. The increase in the bulk modulus over that based
on the average strain assumption can be reversed by incorporating
the anisotropic part of $\overline{\left( \mathbf{A}^{(BA)}\right)
^{-1}}$into the calculation$.$

\section{Conclusion}

We have considered a random aggregate of identical, frictionless,
elastic spheres subjected to an initial confining pressure
followed by a general increment in strain. We have incorporated
the deviation from the average of the difference in the
displacement of a typical pair of particles into the contact
force. We then determined the approximate value of this
fluctuation in terms of statistical measures of the geometry of
the packing and the contact stiffnesses using force equilibrium in
a way suggested by Jenkins (1997). We then made simple but
plausible assumptions regarding the form of the simple and
conditional probabilities that determined the geometry of the
packing and used these to carry out the averages. Of particular
interest was the quantity $\overline{\left(
\mathbf{A}^{(BA)}\right)  ^{-1}\mathbf{J}^{(BA)}}$. This involved
the product of averages and the average of the product of
fluctuations. We found that the product of the averages provides
$46\%$ of the reduction of the shear modulus and the average of
the product provides the remaining $24\%$. Other sources of a
reduction in moduli were not considered.

For example, Trendadue (2001), apparently following Jenkins
(1997), carries out a similar calculation of the bulk and shear
moduli for elastic, frictional spheres and includes a reduction in
moduli due to fluctuations during the initial isotropic
compression and a reduction due to the retention of the
anisotropic part of $\overline{\left(  \mathbf{A}^{(BA)}\right)
^{-1}}$. On the other hand, he does not include a contribution
from the product of the averages. Our experience with the computer
simulations indicates that the reduction of the shear modulus
associated with the former two contributions are relatively small,
but that associated with the the latter is significant.

Independently, Paine (1997) introduced fluctuations in
displacement and used the equilibrium equations to determine them
in terms of the packing and stiffness. She then employed computer
simulations, measured the statistical distributions, and
calculated the reduction of the bulk modulus.

We believe that the correct modeling of the statistical
distributions is crucial to the prediction of the mechanical
behavior of granular materials. In our modeling, we have assumed
them to be the most homogeneous possible. This assumption is
plausible and seems to be effective. We have focused on
frictionless particles in an isotropic compressed state, because,
for frictionless particles, the difference in the values of the
shear modulus based on the average stain assumption and those
measured in the numerical simulation is very large. Consequently,
this material provides a good test of any improved theory. We
believe that we have incorporated the essential features of the
correct distributions into such a theory and have established an
appropriate basis for the extension of the theory to frictional
particles and more complicated states of stress.

Finally, it is natural to question why the average strain
assumption provides a relative good approximation to the bulk
modulus and a relatively poor approximation to the shear modulus.
An indication of why this is so can be obtained by considering a
behavior of a simple model neighborhood of the sphere $A$ in which
sphere $B$ is on the polar axis in the upper hemisphere and three
other spheres in the lower hemisphere are distributed
symmetrically about the polar axis at an angle
$\phi=\cos^{-1}(1/3)$ below the equator. Then, when this
neighborhood is subjected to a isotropic compression, the relative
displacement of the particles $A$ and $B$ is given by the average
strain and there is no fluctuation. However, when the neighborhood
is subjected to a deviatoric strain, there must be a deviation
from the displacements associated with the average strain, whose
direction and magnitude varies with the orientation of the
neighborhood with respect to the principal axes of the strain, in
order for the pair to be in equilibrium. That is, a typical
neighborhood is closer to being equilibrated by a average
isotropic strain than by an average deviatoric strain.

\section{Acknowledgement}

This research was supported, in part, by CNR AGENZIA 2000 and the
Department of Energy.

\appendix

\section{\textbf{Single Particle Relaxation}}

Consider a specific particle, labelled $A$, which we take to be
centered at the origin. It has contacts with particles centered at
$\mathbf{d}^{(nA)}$, $n=1,2,..N^{(A)}$. Assuming that one of the
particle centers is displaced by an increment
$\mathbf{\dot{u}}^{(nA)}$ the form of equation (\ref{Force}) is
\begin{equation}
\dot{F}_{i}^{(nA)}=K_{N}\left(
\hat{d_{k}}^{(nA)}\dot{u}_{k}^{(nA)}\right)
\hat{d_{i}}^{(nA)}, \label{D1}%
\end{equation}
where $K_{N}$ is given by
\[
K_{N}=\frac{\mu d^{1/2}}{1-\nu}\delta^{1/2}%
\]
with $\delta$ given by equation (\ref{delta}).

As written, the total force on the specific particle, due to the
sum of all the incremental contact forces is not zero:
\[
\dot{F}_{i}^{(A)}\equiv\sum_{n=1}^{N^{(A)}}\dot{F}_{i}^{(nA)}\neq0.
\]
Accordingly, that particle will move to a new incremental
position, $\mathbf{\dot{X}}^{(A)}$. The generalization of equation
(\ref{D1}) that takes into account the new position and
orientation is
\[
\dot{F}_{i}^{(nA)}=K_{N}\left[  \hat{d_{k}}^{(nA)}\left(  \dot{u}_{k}%
^{(nA)}-\dot{X}_{k}^{(A)}\right)  \right]  \hat{d_{i}}^{(nA)},
\]
Now, the requirement that the particle is in equilibrium with its
contact
forces gives three linear equations in the three unknowns $\mathbf{\dot{X}%
}^{(A)}$. It is straightforward to solve these equations
numerically.

Having determined the new equilibrium position and orientation,
one can show that the total increment in work done by the
increment in contact forces on particle $A$ is simply
\begin{equation}
\dot{W}^{(A)}=\frac{1}{2}\left[  K_{N}\sum_{n=1}^{N^{(A)}}(\hat{d_{i}}%
^{(nA)}\dot{u}_{i}^{(nA)})^{2}-\dot{F}_{i}^{(A)}\dot{X}_{i}^{(A)}\right]
,
\label{D7}%
\end{equation}
where $\mathbf{X}^{(A)}$ is determined as described above. In
order to calculate $\dot{W}^{(A)}$ we make the average strain
assumption that the displacement at the contact point is simply
related to the macroscopic strain
\begin{equation}
\dot{u}_{i}^{(nA)}=\dot{E}_{ij}d_{j}^{(nA)}. \label{D8}%
\end{equation}
Because we know the exact positions of each center,
$\mathbf{d}^{(nA)}$, from the simulations, we are able to evaluate
equation (\ref{D7}) for each particle in the ensemble.

We now evaluate the elastic moduli by setting the total
deformation in the contacts equal to the macroscopic strain
energy:
\begin{equation}
\frac{1}{V}\sum_{A=1}^{N}\dot{W}^{(A)}=\frac{1}{2}\left[  \left(
\bar{\kappa
}-\frac{2}{3}\bar{\mu}\right)  (\dot{E}_{ll})^{2}+2\bar{\mu}\dot{E}_{ij}%
\dot{E}_{ij}\right]  , \label{D9}%
\end{equation}
where the sum is taken over all particles in the computational
unit cell of volume $V$. The left-hand side of equation (\ref{D9})
can be evaluated once for a pure compression and once for a simple
shear in order to deduce the values of $\bar{\kappa}$ and
$\bar{\mu}$. The point of the exercise is to investigate the
extent to which relaxation, at the single particle level, can
explain the large reduction of the shear modulus relative to the
prediction of the average strain approximation.

If, in equations (\ref{D7}) and (\ref{D8}), we assume there is no
relaxation ($\mathbf{{\dot{X}}}^{(A)}\mathbf{=0}$), and if we
replace the sum over contacts by an integral over a presumed
uniform distribution of contact directions, we reproduce the
average strain theory, equations (\ref{mu}) and (\ref{lambda}), as
detailed in Norris and Johnson (1997).

\section{$z_{\alpha|\beta}(\mathbf{\alpha},\mathbf{\beta)}$}

Here, we determine the appropriate form of the distribution
function $z_{\alpha|\beta}(\mathbf{\alpha},\mathbf{\beta)}$ for
the intervals of polar angles $\theta_{\alpha}$ and
$\theta_{\beta}$ indicated in (\ref{cond_equat}), given that
particle $B$ is on the pole. We assume that the distribution
$z_{\alpha|\beta}(\mathbf{\alpha},\mathbf{\beta)}$ is independent
of the circumferential angle $\phi_{\alpha}$ and that there are
the same number of $\alpha$ contacts in each infinitesimal
circumferential strip about the pole whether or not the particle
$\beta$ intrudes on the strip. This is the simplest kind of
homogeneity that applies in this case.

We first consider the range of polar angles in the first integral:%
\[
\pi/3\leq\theta_{\beta}\leq2\pi/3\text{ and }\pi/3\leq\theta_{\alpha}%
\leq\theta_{\beta}+\pi/3.\text{ }%
\]
In the subinterval $\pi/3\leq\theta_{\beta}\leq\pi/2$,
$g(\mathbf{\beta
})=\left[  (k/2)-1\right]  /\pi=(k-2)/2\pi$. For $\pi/3\leq\theta_{\alpha}%
\leq\pi/2$, the expected average number of $\alpha$ particles in
the upper hemisphere, given the presence of $\beta$ there, is
$(k/2)-2$, so
\begin{align*}
\frac{k-4}{2}  &  =\int_{\frac{\pi}{3}}^{\frac{\pi}{2}}\int_{\Phi
(\theta_{\alpha})}^{2\pi-\Phi(\theta_{\alpha})}z_{\alpha|\beta}(\mathbf{\alpha
},\mathbf{\beta})\sin\theta_{\alpha}d\phi_{\alpha}d\theta_{\alpha}\\
&
=2\int_{\frac{\pi}{3}}^{\frac{\pi}{2}}z_{\alpha|\beta}(\theta_{\alpha
})\left[  \pi-\Phi(\theta_{\alpha})\right]
\sin\theta_{\alpha}d\theta _{\alpha}.
\end{align*}
If $z_{\alpha|\beta}(\theta_{\alpha})\left[
\pi-\Phi(\theta_{\alpha})\right] $ is to be constant, then
\[
z_{\alpha|\beta}(\mathbf{\alpha},\mathbf{\beta})=\frac{k-4}{2(\pi-\Phi)}.
\]
When $\pi/2\leq\theta_{\alpha}\leq\theta_{\beta}+\pi/3$, the
average number $k_{1}$ of $\alpha$ particles in this strip, given
the presence of $\beta$ in the upper hemisphere, is obtained from
the proportion
\[
k/2:2\pi=k_{1}:\int_{0}^{2\pi}\int_{\frac{\pi}{2}}^{\theta_{\beta}+\frac{\pi
}{3}}\sin\theta_{\alpha}d\theta_{\alpha}d\phi_{\alpha}%
\]
as
\[
k_{1}=-\frac{k}{2}\cos\left(  \theta_{\beta}+\frac{\pi}{3}\right)
.
\]
Then,
\[
-\frac{k}{2}\cos\left(  \theta_{\beta}+\frac{\pi}{3}\right)
=\int_{\frac{\pi
}{2}}^{\theta_{\beta}+\frac{\pi}{3}}\int_{\Phi(\theta_{\alpha})}^{2\pi
-\Phi(\theta_{\alpha})}z_{\alpha|\beta}(\mathbf{\alpha},\mathbf{\beta}%
)\sin\theta_{\alpha}d\phi_{\alpha}d\theta_{\alpha};
\]
and
\[
z_{\alpha|\beta}(\mathbf{\alpha},\mathbf{\beta})=\frac{k}{4(\pi-\Phi)}.
\]
In the subinterval $\pi/2\leq\theta_{\beta}\leq2\pi/3$,
$g(\mathbf{\beta })=(k/2)/2\pi=k/4\pi$. For
$\pi/3\leq\theta_{\alpha}\leq\pi/2$, the average number of
$\alpha$ particles in the upper hemisphere, given the presence
$\beta$ in the lower hemisphere, is $(k/2)-1$, so
\[
\frac{k-2}{2}=\int_{\frac{\pi}{3}}^{\frac{\pi}{2}}\int_{\Phi(\theta_{\alpha}%
)}^{2\pi-\Phi(\theta_{\alpha})}z_{\alpha|\beta}(\mathbf{\alpha},\mathbf{\beta
})\sin\theta_{\alpha}d\theta_{\alpha}d\phi_{\alpha},
\]
and
\[
z_{\alpha|\beta}(\mathbf{\alpha},\mathbf{\beta})=\frac{k-2}{2(\pi-\Phi)}.
\]
For $\pi/2\leq\theta_{\alpha}\leq2\pi/3$, the average number of
$\alpha$ particles in this strip is $(k/4)-1$, so
\[
\frac{k-4}{4}=\int_{\frac{\pi}{2}}^{\frac{2\pi}{3}}\int_{\Phi(\theta_{\alpha
})}^{2\pi-\Phi(\theta_{\alpha})}z_{\alpha|\beta}(\mathbf{\alpha}%
,\mathbf{\beta})\sin\theta_{\alpha}d\phi_{\alpha}d\theta_{\alpha},
\]
and\textbf{\ }
\[
z_{\alpha|\beta}(\mathbf{\alpha},\mathbf{\beta})=\frac{k-4}{4(\pi-\Phi)}.
\]
Finally, for $2\pi/3\leq\theta_{\alpha}\leq\theta_{\beta}+\pi/3$,
\[
k/2:2\pi=k_{2}:\int_{0}^{2\pi}\int_{\frac{2\pi}{3}}^{\theta_{\beta}+\frac{\pi
}{3}}\sin\theta_{\alpha}d\theta_{\alpha}d\phi_{\alpha},
\]
so
\[
k_{2}=-\frac{k}{2}\left[  \cos\left(
\theta_{\beta}+\frac{\pi}{3}\right) +\frac{1}{2}\right]  ,
\]%
\[
-\frac{k}{2}\left[  \cos\left(
\theta_{\beta}+\frac{\pi}{3}\right)  +\frac {1}{2}\right]
=\int_{\frac{2\pi}{3}}^{\theta_{\beta}+\frac{\pi}{3}}\int
_{\Phi(\theta_{\alpha})}^{2\pi-\Phi(\theta_{\alpha})}z_{\alpha|\beta
}(\mathbf{\alpha},\mathbf{\beta})\sin\theta_{\alpha}d\phi_{\alpha}%
d\theta_{\alpha},
\]
and
\[
z_{\alpha|\beta}(\mathbf{\alpha},\mathbf{\beta})=\frac{k}{4(\pi-\Phi)}.
\]

The range of polar angles in second integral is
\[
\pi/3\leq\theta_{\beta}\leq2\pi/3\text{ and
}\theta_{\beta}+\pi/3\leq \theta_{\alpha}\leq\pi.
\]
In the subinterval $\pi/3\leq\theta_{\beta}\leq\pi/2$,
$g(\mathbf{\beta })=(k-2)/2\pi;$ and in the subinterval
$\pi/2\leq\theta_{\beta}\leq2\pi/3$, $g(\mathbf{\beta})=k/4\pi$.
For $\theta_{\beta}+\pi/3\leq\theta_{\alpha}\leq$ $\pi$, given
$\beta$ in the first subinterval, the average number $k_{3}$ of
$\alpha$ particles in the strip is determined by
\[
k/2:2\pi=k_{3}:\int_{0}^{2\pi}\int_{\theta_{\beta}+\frac{\pi}{3}}^{\pi}%
\sin\theta_{\alpha}d\theta_{\alpha}d\phi_{\alpha}%
\]
as
\[
k_{3}=\frac{k}{2}\left[  1+\cos\left(
\theta_{\beta}+\frac{\pi}{3}\right) \right]  .
\]
Then
\[
\frac{k}{2}\left[  1+\cos\left(
\theta_{\beta}+\frac{\pi}{3}\right)  \right]
=\int_{\theta_{\beta}+\frac{\pi}{3}}^{\pi}\int_{0}^{2\pi}z_{\alpha|\beta
}(\mathbf{\alpha},\mathbf{\beta})\sin\theta_{\alpha}d\phi_{\alpha}%
d\theta_{\alpha},
\]
and
\[
z_{\alpha|\beta}(\mathbf{\alpha},\mathbf{\beta})=\frac{k}{4\pi}.
\]
Then for $\theta_{\beta}+\pi/3\leq\theta_{\alpha}\leq$ $\pi$ and
$\beta$ in the second subinterval, the distribution function is
the same.

The range of polar angles in third integral is
\[
2\pi/3\leq\theta_{\beta}\leq\pi\text{ and
}\pi/3\leq\theta_{\alpha}\leq \theta_{\beta}-\pi/3.
\]
In the subinterval $2\pi/3\leq\theta_{\beta}\leq5\pi/6$,
$g(\mathbf{\beta })=k/4\pi$ and the average number $k_{4}$ of
$\alpha$ particles in the strip
$\pi/3\leq\theta_{\alpha}\leq\theta_{\beta}-\pi/3$ is determined
by
\[
\left(  k-2\right)  /2:\pi=k_{4}:\int_{0}^{2\pi}\int_{\frac{\pi}{3}}%
^{\theta_{\beta}-\frac{\pi}{3}}\sin\theta_{\alpha}d\theta_{\alpha}%
d\phi_{\alpha}%
\]
as
\[
k_{4}=\left(  k-2\right)  \left[  \frac{1}{2}-\cos\left(  \theta_{\beta}%
-\frac{\pi}{3}\right)  \right]  .
\]
Then
\[
\left(  k-2\right)  \left[  \frac{1}{2}-\cos\left(
\theta_{\beta}-\frac{\pi
}{3}\right)  \right]  =\int_{\frac{\pi}{3}}^{\theta_{\beta}-\frac{\pi}{3}}%
\int_{0}^{2\pi}z_{\alpha|\beta}(\mathbf{\alpha},\mathbf{\beta})\sin
\theta_{\alpha}d\phi_{\alpha}d\theta_{\alpha},
\]
and
\[
z_{\alpha|\beta}(\mathbf{\alpha},\mathbf{\beta})=\frac{\left(
k-2\right) }{2\pi}.
\]
In the subinterval $5\pi/6\leq\theta_{\beta}\leq\pi$,
$g(\mathbf{\beta })=k/4\pi.$ For
$\pi/3\leq\theta_{\alpha}\leq\pi/2$, the average number of
$\alpha$ particles in the upper hemisphere, given the presence of
$\beta$ in the lower hemisphere, is $(k/2)-1$, so
\[
z_{\alpha|\beta}(\mathbf{\alpha},\mathbf{\beta})=\frac{k-2}{2\pi}.
\]
For $\pi/2\leq\theta_{\alpha}\leq\theta_{\beta}-\pi/3$, the
average number $k_{5}$ of $\alpha$ particles in the strip is
determined by
\[
k/2:2\pi=k_{5}:\int_{0}^{2\pi}\int_{\frac{\pi}{2}}^{\theta_{\beta}-\frac{\pi
}{3}}\sin\theta_{\alpha}d\theta_{\alpha}d\phi_{\alpha}%
\]
as
\[
k_{5}=-\frac{k}{2}\cos\left(  \theta_{\beta}-\frac{\pi}{3}\right)
.
\]
Then
\[
-\frac{k}{2}\cos\left(  \theta_{\beta}-\frac{\pi}{3}\right)
=\int_{\frac{\pi
}{2}}^{\theta_{\beta}-\frac{\pi}{3}}\int_{0}^{2\pi}z_{\alpha|\beta
}(\mathbf{\alpha},\mathbf{\beta})\sin\theta_{\alpha}d\phi_{\alpha}%
d\theta_{\alpha},
\]
and
\[
z_{\alpha|\beta}(\mathbf{\alpha},\mathbf{\beta})=\frac{k}{4\pi}.
\]

The range of polar angles in the fourth integral is
\[
2\pi/3\leq\theta_{\beta}\leq\pi\text{ and
}\theta_{\beta}-\pi/3\leq
\theta_{\alpha}\leq5\pi/3-\theta_{\beta},
\]
where $g(\mathbf{\beta})=k/4\pi$. In the subinterval
$\theta_{\beta}-\pi /3\leq\theta_{\alpha}\leq\pi/2$, with
$2\pi/3\leq\theta_{\beta}\leq5\pi/6$,
\[
\left(  k-2\right)
/2:\pi=k_{6}:\int_{0}^{2\pi}\int_{\theta_{\beta}-\frac
{\pi}{3}}^{\frac{\pi}{2}}\sin\theta_{\alpha}d\theta_{\alpha}d\phi_{\alpha},
\]%
\[
k_{6}=(k-2)\cos\left(  \theta_{\beta}-\frac{\pi}{3}\right)  ;
\]
so
\[
(k-2)\cos\left(  \theta_{\beta}-\frac{\pi}{3}\right)
=\int_{\theta_{\beta
}-\frac{\pi}{3}}^{\frac{\pi}{2}}\int_{\Phi(\theta_{\alpha})}^{2\pi-\Phi
(\theta_{\alpha})}z_{\alpha|\beta}(\mathbf{\alpha},\mathbf{\beta})\sin
\theta_{\alpha}d\phi_{\alpha}d\theta_{\alpha}%
\]
and
\[
z_{\alpha|\beta}(\mathbf{\alpha},\mathbf{\beta})=\frac{k-2}{2(\pi-\Phi)}.
\]

Similarly, from
\[
k/2:2\pi=k_{7}:\int_{0}^{2\pi}\int_{\frac{\pi}{2}}^{\frac{5\pi}{3}%
-\theta_{\beta}}\sin\theta_{\alpha}d\theta_{\alpha}d\phi_{\alpha},
\]%
\[
k_{7}=-\frac{k}{2}\cos\left(  \frac{5\pi}{3}-\theta_{\beta}\right)
;
\]
so
\[
k_{7}=\int_{\frac{\pi}{2}}^{5\pi/3-\theta_{\beta}}\int_{\Phi(\theta_{\alpha}%
)}^{2\pi-\Phi(\theta_{\alpha})}z_{\alpha|\beta}(\mathbf{\alpha},\mathbf{\beta
})\sin\theta_{\alpha}d\phi_{\alpha}d\theta_{\alpha}%
\]
and
\[
z_{\alpha|\beta}(\mathbf{\alpha},\mathbf{\beta})=\frac{k}{4(\pi-\Phi)}.
\]
Finally, for
$\theta_{\beta}-\pi/3\leq\theta_{\alpha}\leq5\pi/3-\theta_{\beta
}$, with $5\pi/6\leq\theta_{\beta}\leq\pi,$ the distribution
function is the same.

\section{Details of the Calculation}

\subsection{$\overline{\mathbf{A}^{(BA)^{\prime}}\mathbf{J}^{(BA)^{\prime}}}$}

We consider first the integrals in equation (\ref{cond_equat}) in
which the
limits of the integration over $\phi_{\alpha}$ depend upon $\Phi$ and define%
\[
Q_{jmn}^{(BA)}\equiv\int_{0}^{2\pi}\beta_{j}\beta_{i}d\phi_{\beta}\int
_{\Phi+\phi_{\beta}}^{2\pi-\Phi+\phi_{\beta}}\alpha_{i}\alpha_{m}\alpha
_{n}d\phi_{\alpha}.
\]
When expressed in terms of the angles, this is%

\begin{align*}
&  Q_{jmn}^{(BA)}=\\
&  \pi\left[
2(\pi-\Phi)\sin^{2}\theta_{\alpha}\cos\theta_{\alpha}\cos
^{2}\theta_{\beta}-\sin\Phi\sin2\theta_{\beta}\sin^{3}\theta_{\alpha}\right]
\delta_{mn}\hat{d}_{j}^{(BA)}\\
&  +\frac{\pi}{2}\left[
2(\pi-\Phi)\sin^{2}\theta_{\alpha}\cos\theta_{\alpha
}\sin^{2}\theta_{\beta}-\sin2\Phi\sin^{2}\theta_{\alpha}\cos\theta_{\alpha
}\sin^{2}\theta_{\beta}\right. \\
&  \left.
-2\sin\Phi\sin\theta_{\alpha}\cos^{2}\theta_{\alpha}\sin
2\theta_{\beta}\right]  \left(  \delta_{mj}\hat{d}_{n}^{(BA)}+\delta_{jn}%
\hat{d}_{m}^{(BA)}\right) \\
&  +\frac{\pi}{2}\left[
4\sin\Phi\sin\theta_{\beta}\cos\theta_{\beta}\sin
^{3}\theta_{\alpha}+8(\pi-\Phi)\cos^{3}\theta_{\alpha}\cos^{2}\theta_{\beta
}\right. \\
&  \left.
-4(\pi-\Phi)\sin^{2}\theta_{\alpha}\cos\theta_{\alpha}+2\sin
2\Phi\sin^{2}\theta_{\alpha}\cos\theta_{\alpha}\sin^{2}\theta_{\beta}\right]
\hat{d}_{m}^{(BA)}\hat{d}_{j}^{(BA)}\hat{d}_{n}^{(BA)}.
\end{align*}
With this, the integral
\[
R_{jmn}^{(BA)}\equiv\int_{\Omega_{\mathbf{\beta}}}\int_{\Omega_{\mathbf{\alpha
}}}g(\mathbf{\beta})z_{\alpha|\beta}(\mathbf{\alpha},\mathbf{\beta})\beta
_{j}\beta_{i}\alpha_{i}\alpha_{m}\alpha_{n}d\Omega_{\alpha}d\Omega_{\beta}%
\]
in equation (\ref{cond_equat}) can be written as%
\begin{align*}
R_{jmn}^{(BA)}  &  =\int_{\frac{\pi}{3}}^{\frac{2\pi}{3}}\int_{\frac{\pi}{3}%
}^{\theta_{\beta}+\frac{\pi}{3}}g(\mathbf{\beta})z_{\alpha|\beta
}(\mathbf{\alpha},\mathbf{\beta})Q_{jmn}^{(BA)}\sin\theta_{a}\sin\theta
_{\beta}d\theta_{\beta}d\theta_{a}\\
&  +\int_{\frac{\pi}{3}}^{\frac{2\pi}{3}}\int_{\theta_{\beta}+\frac{\pi}{3}%
}^{\pi}\int_{0}^{2\pi}\int_{0}^{2\pi}g(\mathbf{\beta})z_{\alpha|\beta
}(\mathbf{\alpha},\mathbf{\beta})\beta_{j}\beta_{i}\alpha_{i}\alpha_{m}%
\alpha_{n}d\Omega_{\alpha}d\Omega_{\beta}\\
&
+\int_{\frac{2\pi}{3}}^{\pi}\int_{\frac{\pi}{3}}^{\theta_{\beta}-\frac{\pi
}{3}}\int_{0}^{2\pi}\int_{0}^{2\pi}g(\mathbf{\beta})z_{\alpha|\beta
}(\mathbf{\alpha},\mathbf{\beta})\beta_{j}\beta_{i}\alpha_{i}\alpha_{m}%
\alpha_{n}d\Omega_{\alpha}d\Omega_{\beta}\\
&
+\int_{\frac{2\pi}{3}}^{\pi}\int_{\theta_{\beta}-\frac{\pi}{3}}^{\frac
{5\pi}{3}-\theta_{\beta}}g(\mathbf{\beta})z_{\alpha|\beta}(\mathbf{\alpha
},\mathbf{\beta})Q_{jmn}^{(BA)}\sin\theta_{a}\sin\theta_{\beta}d\theta_{\beta
}d\theta_{a},
\end{align*}
We use the appropriate distribution function for each range of
variation of
the polar angles and distinguish the integrals that involve $\mathbf{Q}%
^{(BA)}$ from those not.

Then, for example,
\begin{align*}
&
\int_{\frac{\pi}{3}}^{\frac{2\pi}{3}}\int_{\frac{\pi}{3}}^{\theta_{\beta
}+\frac{\pi}{3}}g(\mathbf{\beta})z_{\alpha|\beta}(\mathbf{\alpha
},\mathbf{\beta})Q_{jmn}^{(BA)}\sin\theta_{a}\sin\theta_{\beta}d\theta_{\beta
}d\theta_{a}\\
&
=\frac{(k-2)}{4\pi}(k-4)\int_{\frac{\pi}{3}}^{\frac{\pi}{2}}\int_{\frac
{\pi}{3}}^{\frac{\pi}{2}}\frac{Q_{jmn}^{(BA)}}{(\pi-\Phi)}\sin\theta_{\beta
}\sin\theta_{\alpha}d\theta_{\beta}d\theta_{\alpha}\\
&  +\frac{(k-2)}{8\pi}k\int_{\frac{\pi}{3}}^{\frac{\pi}{2}}\int_{\frac{\pi}%
{2}}^{\theta_{\beta}+\frac{\pi}{3}}\frac{Q_{jmn}^{(BA)}}{(\pi-\Phi)}\sin
\theta_{\beta}\sin\theta_{\alpha}d\theta_{\beta}d\theta_{\alpha}\\
&
+\frac{k}{8\pi}(k-2)\int_{\frac{\pi}{2}}^{\frac{2\pi}{3}}\int_{\frac{\pi
}{3}}^{\frac{\pi}{2}}\frac{Q_{jmn}^{(BA)}}{(\pi-\Phi)}\sin\theta_{\beta}%
\sin\theta_{\alpha}d\theta_{\beta}d\theta_{\alpha}\\
&
+\frac{k}{16\pi}(k-4)\int_{\frac{\pi}{2}}^{\frac{2\pi}{3}}\int_{\frac{\pi
}{2}}^{\frac{2\pi}{3}}\frac{Q_{jmn}^{(BA)}}{(\pi-\Phi)}\sin\theta_{\beta}%
\sin\theta_{\alpha}d\theta_{\beta}d\theta_{\alpha}\\
&
+\frac{k^{2}}{16\pi}\int_{\frac{\pi}{2}}^{\frac{2\pi}{3}}\int_{\frac{2\pi
}{3}}^{\theta_{\beta}+\frac{\pi}{3}}\frac{Q_{jmn}^{(BA)}}{(\pi-\Phi)}%
\sin\theta_{\beta}\sin\theta_{\alpha}d\theta_{\beta}d\theta_{\alpha}.
\end{align*}
Upon evaluation the integrals over $\theta_{\beta}$ and
$\theta_{\alpha}$ numerically, we obtain
\begin{align*}
&
\int_{\frac{\pi}{3}}^{\frac{2\pi}{3}}\int_{\frac{\pi}{3}}^{\theta_{\beta
}+\frac{\pi}{3}}g(\mathbf{\beta})z_{\alpha|\beta}(\mathbf{\alpha
},\mathbf{\beta})Q_{jmn}^{(BA)}\sin\theta_{a}\sin\theta_{\beta}d\theta_{\beta
}d\theta_{\alpha}\\
&  =\left[  -0.13\frac{(k-2)}{4\pi}(k-4)+0.57\frac{k(k-2)}{8\pi}%
-0.38\frac{k(k-2)}{8\pi}\right. \\
&  \left.
+0.13\frac{k}{16\pi}(k-4)+0.19\frac{k^{2}}{16\pi}\right]  \hat
{d}_{n}^{(BA)}\hat{d}_{j}^{(BA)}\hat{d}_{m}^{(BA)}\\
&  +\left[
0.11\frac{(k-2)}{4\pi}(k-4)-0.23\frac{(k-2)}{8\pi}k+0.13\frac
{k}{8\pi}(k-2)\right. \\
&  \left.
-0.11\frac{k}{16\pi}(k-4)-0.11\frac{k^{2}}{16\pi}\right]  \left(
\delta_{mj}\hat{d}_{n}^{(BA)}+\delta_{jn}\hat{d}_{m}^{(BA)}\right) \\
&  +\left[
-0.11\frac{(k-2)}{4\pi}(k-4)-0.16\frac{(k-2)}{8\pi}k+0.14\frac
{k}{8\pi}(k-2)\right. \\
&  \left.
+0.11\frac{k}{16\pi}(k-4)+0.01\frac{k^{2}}{16\pi}\right]
\delta_{mn}\hat{d}_{j}^{(BA)}%
\end{align*}
In the same way,%
\begin{align*}
&
\int_{\frac{2\pi}{3}}^{\pi}\int_{\theta_{\beta}-\frac{\pi}{3}}^{\frac{5\pi
}{3}-\theta_{\beta}}g(\mathbf{\beta})z_{\alpha|\beta}(\mathbf{\alpha
},\mathbf{\beta})Q_{jmn}^{(BA)}\sin\theta_{a}\sin\theta_{\beta}d\theta_{\beta
}d\theta_{a}\\
&  =\frac{k(k-2)}{8\pi}\int_{\frac{2\pi}{3}}^{\frac{5\pi}{6}}\int
_{\theta_{\beta}-\frac{\pi}{3}}^{\frac{\pi}{2}}\frac{Q_{jmn}^{(BA)}}{(\pi
-\Phi)}\sin\theta_{\alpha}\sin\theta_{\beta}d\theta_{\beta}d\theta_{\alpha}\\
&
+\frac{k^{2}}{16\pi}\int_{\frac{2\pi}{3}}^{\frac{5\pi}{6}}\int_{\frac{\pi
}{2}}^{\frac{5\pi}{3}-\theta_{\beta}}\frac{Q_{jmn}^{(BA)}}{(\pi-\Phi)}%
\sin\theta_{\alpha}\sin\theta_{\beta}d\theta_{\beta}d\theta_{\alpha}\\
&  +\frac{k^{2}}{16\pi}\int_{\frac{5\pi}{6}}^{\pi}\int_{\theta_{\beta}%
-\frac{\pi}{3}}^{\frac{5\pi}{3}-\theta_{\beta}}\frac{Q_{jmn}^{(BA)}}{(\pi
-\Phi)}\sin\theta_{\alpha}\sin\theta_{\beta}d\theta_{\beta}d\theta_{\alpha}%
\end{align*}%
\begin{align*}
&  =\frac{k(k-2)}{8\pi}\left[  -0.15\hat{d}_{m}^{(BA)}\hat{d}_{j}^{(BA)}%
\hat{d}_{n}^{(BA)}+0.03\left(  \delta_{mj}\hat{d}_{n}^{(BA)}+\delta_{jn}%
\hat{d}_{m}^{(BA)}\right)  \right. \\
&  \left.  +0.10\delta_{mn}\hat{d}_{j}^{(BA)}\right] \\
&  +\frac{k^{2}}{16\pi}\left(  -0.23\hat{d}_{m}^{(BA)}\hat{d}_{j}^{(BA)}%
\hat{d}_{n}^{(BA)}-0.05\delta_{mn}\hat{d}_{j}^{(BA)}\right)  .
\end{align*}

In the integrals that are independent of $\Phi$, we can easily
carry out the integration of
\[
P_{jmn}^{(BA)}\equiv\frac{1}{2\pi}\int_{0}^{2\pi}\beta_{j}\beta_{i}%
d\phi_{\beta}\frac{1}{2\pi}\int_{0}^{2\pi}\alpha_{i}\alpha_{m}\alpha_{n}%
d\phi_{\alpha}:
\]%
\begin{align*}
P_{jmn}^{(BA)}  &  =\hat{d}_{m}^{(BA)}\hat{d}_{j}^{(BA)}\hat{d}_{n}%
^{(BA)}\left[  \cos^{2}\theta_{\beta}\left(
\cos^{3}\theta_{\alpha}-\frac
{3}{2}\cos\theta_{\alpha}\sin^{2}\theta_{\alpha}\right)  \right. \\
&  \left.  -\frac{1}{2}\sin^{2}\theta_{\beta}\cos\theta_{\alpha}\sin^{2}%
\theta_{\alpha}+\cos^{2}\theta_{\beta}\cos\theta_{\alpha}\sin^{2}%
\theta_{\alpha}\right] \\
&
+\frac{1}{2}\cos^{2}\theta_{\beta}\cos\theta_{\alpha}\sin^{2}\theta
_{\alpha}\hat{d}_{j}^{(BA)}\delta_{mn}\\
&
+\frac{1}{4}\sin^{2}\theta_{\beta}\cos\theta_{\alpha}\sin^{2}\theta
_{\alpha}\left(  \delta_{jm}\hat{d}_{n}^{(BA)}+\delta_{nj}\hat{d}_{m}%
^{(BA)}\right)  .
\end{align*}
With this,
\begin{align*}
&  \int_{\frac{\pi}{3}}^{\frac{2\pi}{3}}\int_{\theta_{\beta}+\frac{\pi}{3}%
}^{\pi}\int_{0}^{2\pi}\int_{0}^{2\pi}g(\mathbf{\beta})z_{\alpha|\beta
}(\mathbf{\alpha},\mathbf{\beta})\beta_{j}\beta_{i}\alpha_{i}\alpha_{m}%
\alpha_{n}d\Omega_{\alpha}d\Omega_{\beta}\\
&
=\frac{k(k-2)}{2}\int_{\frac{\pi}{3}}^{\frac{\pi}{2}}\int_{\theta_{\beta
}+\frac{\pi}{3}}^{\pi}P_{jmn}^{(BA)}\sin\theta_{a}\sin\theta_{\beta}%
d\theta_{\beta}d\theta_{\beta}\\
&
+\frac{k^{2}}{4}\int_{\frac{\pi}{2}}^{\frac{2\pi}{3}}\int_{\theta_{\beta
}+\frac{\pi}{3}}^{\pi}P_{jmn}^{(BA)}\sin\theta_{a}\sin\theta_{\beta}%
d\theta_{\beta}d\theta_{a}%
\end{align*}%
\begin{align*}
&  =\frac{k(k-2)}{8\pi}\left[
0.10\hat{d}_{m}^{(BA)}\hat{d}_{j}^{(BA)}\hat
{d}_{n}^{(BA)}-0.09\left(
\delta_{jm}\hat{d}_{n}^{(BA)}+\delta_{jn}\hat
{d}_{m}^{(BA)}\right)  \right. \\
&  \left.  -0.03\hat{d}_{j}^{(BA)}\delta_{mn}\right] \\
&  +\frac{k^{2}}{16\pi}\left[
0.03\hat{d}_{i}^{(BA)}\hat{d}_{j}^{(BA)}\hat
{d}_{p}^{(BA)}-0.01\left(
\delta_{ji}\hat{d}_{p}^{(BA)}+\delta_{ip}\hat
{d}_{j}^{(BA)}\right)  \right]
\end{align*}
and
\begin{align*}
&
\int_{\frac{2\pi}{3}}^{\pi}\int_{\frac{\pi}{3}}^{\theta_{\beta}-\frac{\pi
}{3}}\int_{0}^{2\pi}\int_{0}^{2\pi}g(\mathbf{\beta})z_{\alpha|\beta
}(\mathbf{\alpha},\mathbf{\beta})\beta_{j}\beta_{i}\alpha_{i}\alpha_{m}%
\alpha_{n}d\Omega_{\alpha}d\Omega_{\beta}\\
&  =\frac{k(k-2)}{2}\int_{\frac{2\pi}{3}}^{\frac{5\pi}{6}}\int_{\frac{\pi}{3}%
}^{\theta_{\beta}-\frac{\pi}{3}}P_{jmn}^{(BA)}\sin\theta_{\beta}\sin
\theta_{\alpha}d\theta_{\beta}d\theta_{\alpha}\\
&
+\frac{k(k-2)}{2}\int_{\frac{5\pi}{6}}^{\pi}\int_{\frac{\pi}{3}}^{\frac
{\pi}{2}}P_{jmn}^{(BA)}\sin\theta_{\beta}\sin\theta_{\alpha}d\theta_{\beta
}d\theta_{\alpha}\\
&  +\frac{k^{2}}{4}\int_{\frac{5\pi}{6}}^{\pi}\int_{\frac{\pi}{2}}%
^{\theta_{\beta}-\frac{\pi}{3}}P_{jmn}^{(BA)}\sin\theta_{\beta}\sin
\theta_{\alpha}d\theta_{\beta}d\theta_{\alpha}%
\end{align*}%
\begin{align*}
&  =\frac{k(k-2)}{8\pi}\left[  -0.12\hat{d}_{m}^{(BA)}\hat{d}_{j}^{(BA)}%
\hat{d}_{n}^{(BA)}+0.03\left(  \delta_{jm}\hat{d}_{n}^{(BA)}+\delta_{jn}%
\hat{d}_{m}^{(BA)}\right)  \right. \\
&  \left.  +0.08\hat{d}_{j}^{(BA)}\delta_{mn}\right] \\
&  +\frac{k(k-2)}{8\pi}\left[  -0.07\hat{d}_{m}^{(BA)}\hat{d}_{j}^{(BA)}%
\hat{d}_{n}^{(BA)}+0.01\left(  \delta_{jm}\hat{d}_{n}^{(BA)}+\delta_{jn}%
\hat{d}_{m}^{(BA)}\right)  \right. \\
&  \left.  +0.08\hat{d}_{j}^{(BA)}\delta_{mn}\right] \\
&  +\frac{k^{2}}{16\pi}\left(
0.02\hat{d}_{m}^{(BA)}\hat{d}_{j}^{(BA)}\hat
{d}_{n}^{(BA)}-0.02\hat{d}_{j}^{(BA)}\delta_{mn}\right)  .
\end{align*}

Then%
\begin{align}
16\pi R_{jmn}^{(BA)}  &  =-\left[
0.52(k-2)(k-4)+0.10k(k-2)\right.
\nonumber\\
&  \left.  -0.13k(k-4)-0.01k^{2}\right]  \hat{d}_{j}^{(BA)}\hat{d}_{m}%
^{(BA)}\hat{d}_{n}^{(BA)}\nonumber\\
&  +\left[  0.44(k-2)(k-4)-0.24k(k-2)\right. \nonumber\\
&  \left.  -0.11k(k-4)-0.14k^{2}\right]  \left(  \delta_{jm}\hat{d}_{n}%
^{(BA)}+\delta_{nj}\hat{d}_{m}^{(BA)}\right) \nonumber\\
&  -\left[  0.44(k-2)(k-4)-0.42k(k-2)\right. \nonumber\\
&  \left.  -0.11k(k-4)+0.04k^{2}\right]
\delta_{mn}\hat{d}_{j}^{(BA)}
\label{ARE}%
\end{align}
Finally, we write the first integral in (\ref{FlucProd}) as
\begin{align}
S_{jmn}^{(BA)}  &  \equiv\int_{\Omega_{\beta}}\int_{\Omega_{\mathbf{\alpha}}%
}F(\mathbf{\alpha},\mathbf{\beta})\beta_{j}\beta_{i}\alpha_{i}\alpha_{m}%
\alpha_{n}d\Omega_{\alpha}d\Omega_{\beta}\nonumber\\
&  =\widetilde{\omega}_{1}\hat{d}_{j}^{(BA)}\hat{d}_{m}^{(BA)}\hat{d}%
_{n}^{(BA)}+\omega_{2}\left(
\delta_{jm}\hat{d}_{n}^{(BA)}+\delta_{nj}\hat
{d}_{m}^{(BA)}\right)  +\omega_{2}\delta_{mn}\hat{d}_{j}^{(BA)}\nonumber\\
&  +R_{jmn}^{(BA)} \label{Fifth_part}%
\end{align}

\subsection{\bigskip$\overline{\mathbf{A}^{(BA)^{\prime}}\mathbf{A}%
^{(BA)^{\prime}}}$}

For the calculation of
$\overline{A_{js}^{(BA)\prime}A_{sl}^{(BA)\prime}}$ we introduce
the integral
\[
F_{jl}\equiv\int_{0}^{2\pi}\beta_{j}\beta_{s}d\phi_{\beta}\int_{\Phi
+\phi_{\beta}}^{2\pi-\Phi+\phi_{\beta}}\alpha_{s}\alpha_{l}d\phi_{\alpha}.
\]
When expressed in terms of the angles,%
\begin{align*}
F_{jl}  &  =-\frac{\pi}{2}\left[
\sin2\Phi\sin^{2}\theta_{\alpha}\sin
^{2}\theta_{\beta}+\sin\Phi\sin2\theta_{\alpha}\sin2\theta_{\beta}\right. \\
&  \left.
-2(\pi-\Phi)\sin^{2}\theta_{\alpha}\sin^{2}\theta_{\beta} \right]
\delta_{jl}\\
&  +\frac{\pi}{2}\left[
\sin2\Phi\sin^{2}\theta_{\alpha}\sin^{2}\theta
_{\beta}-\sin\Phi\sin2\theta_{\alpha}\sin2\theta_{\beta}\right. \\
&  \left.  -2(\pi-\Phi)\sin^{2}\theta_{\alpha}\sin^{2}\theta_{\beta}%
+8(\pi-\Phi)\cos^{2}\theta_{\alpha}\cos^{2}\theta_{\beta}\right]  \hat{d}%
_{j}\hat{d}_{l}%
\end{align*}
With this, the integral
\[
Y_{jl}^{(BA)}\equiv\int_{\Omega_{\mathbf{\beta}}}\int_{\Omega_{\mathbf{\alpha
}}}F_{jl}^{(BA)}g(\mathbf{\beta})z_{\alpha|\beta}(\mathbf{\alpha
},\mathbf{\beta})\beta_{j}\beta_{s}\alpha_{s}\alpha_{m}d\Omega_{\alpha}%
d\Omega_{\beta}%
\]
in equation (\ref{cond_equat}) can be written as%
\begin{align*}
Y_{jl}^{(BA)}  &  =\int_{\frac{\pi}{3}}^{\frac{2\pi}{3}}\int_{\frac{\pi}{3}%
}^{\theta_{\beta}+\frac{\pi}{3}}F_{jl}^{(BA)}g(\mathbf{\beta})z_{\alpha|\beta
}(\mathbf{\alpha},\mathbf{\beta})\sin\theta_{a}\sin\theta_{\beta}%
d\theta_{\beta}d\theta_{a}\\
&  +\int_{\frac{\pi}{3}}^{\frac{2\pi}{3}}\int_{\theta_{\beta}+\frac{\pi}{3}%
}^{\pi}\int_{0}^{2\pi}\int_{0}^{2\pi}g(\mathbf{\beta})z_{\alpha|\beta
}(\mathbf{\alpha},\mathbf{\beta})\beta_{j}\beta_{s}\alpha_{s}\alpha_{l}%
d\Omega_{\alpha}d\Omega_{\beta}\\
&
+\int_{\frac{2\pi}{3}}^{\pi}\int_{\frac{\pi}{3}}^{\theta_{\beta}-\frac{\pi
}{3}}\int_{0}^{2\pi}\int_{0}^{2\pi}g(\mathbf{\beta})z_{\alpha|\beta
}(\mathbf{\alpha},\mathbf{\beta})\beta_{j}\beta_{s}\alpha_{s}\alpha_{l}%
d\Omega_{\alpha}d\Omega_{\beta}\\
&
+\int_{\frac{2\pi}{3}}^{\pi}\int_{\theta_{\beta}-\frac{\pi}{3}}^{\frac
{5\pi}{3}-\theta_{\beta}}F_{jl}^{(BA)}g(\mathbf{\beta})z_{\alpha|\beta
}(\mathbf{\alpha},\mathbf{\beta})\sin\theta_{a}\sin\theta_{\beta}%
d\theta_{\beta}d\theta_{a},
\end{align*}
Then, for example, the first integral is
\begin{align*}
&
\int_{\frac{\pi}{3}}^{\frac{2\pi}{3}}\int_{\frac{\pi}{3}}^{\theta_{\beta
}+\frac{\pi}{3}}g(\mathbf{\beta})z_{\alpha|\beta}(\mathbf{\alpha
},\mathbf{\beta})F_{jl}^{(BA)}\sin\theta_{a}\sin\theta_{\beta}d\theta_{\beta
}d\theta_{a}\\
&
=\frac{(k-2)}{4\pi}(k-4)\int_{\frac{\pi}{3}}^{\frac{\pi}{2}}\int_{\frac
{\pi}{3}}^{\frac{\pi}{2}}\frac{F_{jl}^{(BA)}}{(\pi-\Phi)}\sin\theta_{\beta
}\sin\theta_{\alpha}d\theta_{\beta}d\theta_{\alpha}\\
&  +\frac{(k-2)}{8\pi}k\int_{\frac{\pi}{3}}^{\frac{\pi}{2}}\int_{\frac{\pi}%
{2}}^{\theta_{\beta}+\frac{\pi}{3}}\frac{F_{jl}^{(BA)}}{(\pi-\Phi)}\sin
\theta_{\beta}\sin\theta_{\alpha}d\theta_{\beta}d\theta_{\alpha}\\
&
+\frac{k}{8\pi}(k-2)\int_{\frac{\pi}{2}}^{\frac{2\pi}{3}}\int_{\frac{\pi
}{3}}^{\frac{\pi}{2}}\frac{F_{jl}^{(BA)}}{(\pi-\Phi)}\sin\theta_{\beta}%
\sin\theta_{\alpha}d\theta_{\beta}d\theta_{\alpha}\\
&
+\frac{k}{16\pi}(k-4)\int_{\frac{\pi}{2}}^{\frac{2\pi}{3}}\int_{\frac{\pi
}{2}}^{\frac{2\pi}{3}}\frac{F_{jl}^{(BA)}}{(\pi-\Phi)}\sin\theta_{\beta}%
\sin\theta_{\alpha}d\theta_{\beta}d\theta_{\alpha}\\
&
+\frac{k^{2}}{16\pi}\int_{\frac{\pi}{2}}^{\frac{2\pi}{3}}\int_{\frac{2\pi
}{3}}^{\theta_{\beta}+\frac{\pi}{3}}\frac{F_{jl}^{(BA)}}{(\pi-\Phi)}\sin
\theta_{\beta}\sin\theta_{\alpha}d\theta_{\beta}d\theta_{\alpha}.
\end{align*}
Upon evaluating the integrals over $\theta_{\alpha}$ and
$\theta_{\beta}$ numerically, we obtain
\begin{align*}
&
\int_{\frac{\pi}{3}}^{\frac{2\pi}{3}}\int_{\pi/3}^{\theta_{\beta}+\frac
{\pi}{3}}g(\mathbf{\beta})z_{\alpha|\beta}(\mathbf{\alpha},\mathbf{\beta
})F_{jl}^{(BA)}\sin\theta_{a}\sin\theta_{\beta}d\theta_{\beta}d\theta_{a}\\
&  =\frac{(k-2)}{4\pi}(k-4)\left(  0.49\delta_{jl}-0.54\hat{d}_{j}^{(BA)}%
\hat{d}_{l}^{(BA)}\right) \\
&  +\frac{(k-2)}{8\pi}k\left(
0.71\delta_{jl}-0.59\hat{d}_{j}^{(BA)}\hat
{d}_{l}^{(BA)}\right) \\
&  +\frac{k}{8\pi}(k-2)\left(
0.55\delta_{jl}-0.47\hat{d}_{j}^{(BA)}\hat
{d}_{l}^{(BA)}\right) \\
&  +\frac{k}{16\pi}(k-4)\left(
0.49\delta_{jl}-0.54\hat{d}_{j}^{(BA)}\hat
{d}_{l}^{(BA)}\right) \\
&  +\frac{k^{2}}{16\pi}\left(
0.17\delta_{jl}-0.16\hat{d}_{j}^{(BA)}\hat {d}_{l}^{(BA)}\right)
.
\end{align*}
Similarly, for the second integral,
\begin{align*}
&
\int_{\frac{2\pi}{3}}^{\pi}\int_{\theta_{\beta}-\frac{\pi}{3}}^{\frac{5\pi
}{3}-\theta_{\beta}}g(\mathbf{\beta})z_{\alpha|\beta}(\mathbf{\alpha
},\mathbf{\beta})F_{jl}^{(BA)}\sin\theta_{\beta}\sin\theta_{\alpha}%
d\theta_{\alpha}d\theta_{\beta}\\
&  =\frac{k(k-2)}{8\pi}\int_{\frac{2\pi}{3}}^{\frac{5\pi}{6}}\int
_{\theta_{\beta}-\frac{\pi}{3}}^{\frac{\pi}{2}}\frac{F_{jl}^{(BA)}\sin
\theta_{\beta}\sin\theta_{\alpha}d\theta_{\alpha}d\theta_{\beta}}{(\pi-\Phi
)}\\
&  +\frac{k^{2}}{16\pi}\int_{\frac{2\pi}{3}}^{5\pi/6}\int_{\frac{\pi}{2}%
}^{\frac{5\pi}{3}-\theta_{\beta}}\frac{F_{jl}^{(BA)}\sin\theta_{\beta}%
\sin\theta_{\alpha}d\theta_{\alpha}d\theta_{\beta}}{(\pi-\Phi)}\\
&  +\frac{k^{2}}{16\pi}\int_{\frac{5\pi}{6}}^{\pi}\int_{\theta_{\beta}%
-\frac{\pi}{3}}^{\frac{5\pi}{3}-\theta_{\beta}}\frac{F_{jl}^{(BA)}\sin
\theta_{\beta}\sin\theta_{\alpha}d\theta_{\alpha}d\theta_{\beta}}{(\pi-\Phi)}%
\end{align*}%
\begin{align*}
&  =\frac{k(k-2)}{8\pi}\left(
0.16\delta_{jl}-0.12\hat{d}_{j}^{(BA)}\hat
{d}_{l}^{(BA)}\right) \\
&  +\frac{k^{2}}{16\pi}\left(
0.14\delta_{jl}+0.08\hat{d}_{j}^{(BA)}\hat
{d}_{l}^{(BA)}\right) \\
&  -\frac{k^{2}}{16\pi}\left(
0.01\delta_{jl}-0.14\hat{d}_{j}^{(BA)}\hat {d}_{l}^{(BA)}\right)
.
\end{align*}
In order to calculate the last two integrals, we first introduce
\begin{align*}
C_{jl}^{(BA)}  &  \equiv\frac{1}{2\pi}\int_{0}^{2\pi}\beta_{j}\beta_{s}%
d\phi_{\beta}\frac{1}{2\pi}\int_{0}^{2\pi}\alpha_{s}\alpha_{l}d\phi_{\alpha}\\
&  =\frac{1}{4}\sin^{2}\theta_{\beta}\sin^{2}\theta_{\alpha}\delta
_{jl}+\left(  \cos^{2}\theta_{\beta}\cos^{2}\theta_{\alpha}-\frac{1}{4}%
\sin^{2}\theta_{\beta}\sin^{2}\theta_{\alpha}\right)  \hat{d}_{j}^{(BA)}%
\hat{d}_{l}^{(BA)}.
\end{align*}
Then,
\begin{align*}
&  \int_{\frac{\pi}{3}}^{\frac{2\pi}{3}}\int_{\theta_{\beta}+\frac{\pi}{3}%
}^{\pi}\int_{0}^{2\pi}\int_{0}^{2\pi}g(\mathbf{\beta})z_{\alpha|\beta
}(\mathbf{\alpha},\mathbf{\beta})\beta_{j}\beta_{s}\alpha_{s}\alpha_{l}%
d\Omega_{\alpha}d\Omega_{\beta}\\
&
=\frac{k(k-2)}{2}\int_{\frac{\pi}{3}}^{\frac{\pi}{2}}\int_{\theta_{\beta
}+\frac{\pi}{3}}^{\pi}C_{jl}^{(BA)}\sin\theta_{a}\sin\theta_{\beta}%
d\theta_{\beta}d\theta_{\beta}\\
&
+\frac{k^{2}}{4}\int_{\frac{\pi}{2}}^{\frac{2\pi}{3}}\int_{\theta_{\beta
}+\frac{\pi}{3}}^{\pi}C_{jl}^{(BA)}\sin\theta_{a}\sin\theta_{\beta}%
d\theta_{\beta}d\theta_{a}\\
&  =\frac{k(k-2)}{8\pi}\left(
0.12\delta_{jl}+0.01\hat{d}_{j}^{(BA)}\hat {d}_{l}^{(BA)}\right)
+\frac{k^{2}}{16\pi}0.01\delta_{jl}.
\end{align*}
and
\begin{align*}
&
\int_{\frac{2\pi}{3}}^{\pi}\int_{\frac{\pi}{3}}^{\theta_{\beta}-\frac{\pi
}{3}}g(\mathbf{\beta})z_{\alpha|\beta}(\mathbf{\alpha},\mathbf{\beta}%
)C_{jl}^{(BA)}\sin\theta_{\beta}\sin\theta_{\alpha}d\theta_{\alpha}%
d\theta_{\beta}\\
&  =\frac{k(k-2)}{2}\int_{\frac{2\pi}{3}}^{\frac{5\pi}{6}}\int_{\frac{\pi}{3}%
}^{\theta_{\beta}-\frac{\pi}{3}}C_{jl}^{(BA)}\sin\theta_{\beta}\sin
\theta_{\alpha}d\theta_{\alpha}d\theta_{\alpha}\\
&
+\frac{k(k-2)}{2}\int_{\frac{5\pi}{6}}^{\pi}\int_{\frac{\pi}{3}}^{\frac
{\pi}{2}}C_{jl}^{(BA)}\sin\theta_{\beta}\sin\theta_{\alpha}d\theta_{\alpha
}d\theta_{\beta}\\
&  +\frac{k^{2}}{4}\int_{\frac{5\pi}{6}}^{\pi}\int_{\frac{\pi}{2}}%
^{\theta_{\beta}-\frac{\pi}{3}}C_{jl}^{(BA)}\sin\theta_{\beta}\sin
\theta_{\alpha}d\theta_{\alpha}d\theta_{\beta}%
\end{align*}%
\begin{align*}
&  =\frac{k(k-2)}{8\pi}\left(
0.09\delta_{jl}-0.02\hat{d}_{j}^{(BA)}\hat
{d}_{l}^{(BA)}\right) \\
&  +\frac{k(k-2)}{8\pi}\left(
0.02\delta_{jl}+0.04\hat{d}_{j}^{(BA)}\hat {d}_{l}^{(BA)}\right)
+\frac{k^{2}}{16\pi}0.01\delta_{jl}.
\end{align*}
So,%
\begin{align}
&  16\pi Y_{jl}^{(BA)}\nonumber\\
&  =[1.96(k-2)(k-4)+3.30k(k-2)+0.49k(k-4)+0.32k^{2}]\delta_{jl}\nonumber\\
&  -\left[  2.16(k-2)(k-4)+2.30k(k-2)+0.54k(k-4)-0.06k^{2}\right]  \hat{d}%
_{j}^{(BA)}\hat{d}_{l}^{(BA)}.\nonumber\\
&  \label{why}%
\end{align}
Finally, we write the first integral in (\ref{Fluct_ayes}) as
\begin{align}
H_{jl}^{(BA)}  &  =\int_{\Omega_{\beta}}\int_{\Omega_{\mathbf{\alpha}}%
}F(\mathbf{\alpha},\mathbf{\beta})\beta_{j}\beta_{i}\alpha_{i}\alpha
_{l}d\Omega_{\alpha}d\Omega_{\beta}\label{fourth_part}\\
&  =Y_{jl}^{(BA)}+\alpha_{1}\delta_{jl}+\widetilde{\alpha}_{2}\hat{d}%
_{j}^{(BA)}\hat{d}_{l}^{(BA)}.\nonumber
\end{align}

\subsection{Final evaluation}

In order to compare the values of the elastic moduli obtained by
numerical simulation with what predicted by the theory, we have to
evaluate all the previous quantities. These are functions of the
coordination number, which is assumed to be $k=6.07.$ Therefore,
with equation (\ref{ARE}), we can write
\[
R_{jmn}^{(BA)}=\chi_{1}\hat{d}_{m}^{(BA)}\hat{d}_{j}^{(BA)}\hat{d}_{n}%
^{(BA)}+\chi_{2}\left(  \delta_{mj}\hat{d}_{n}^{(BA)}+\delta_{jn}\hat{d}%
_{m}^{(BA)}\right)  +\chi_{3}\delta_{mn}\hat{d}_{j}^{(BA)},
\]
where $\chi_{1}=-0.10,$ $\chi_{2}=-0.17,$ and $\chi_{3}=0.13$ and,
with equation (\ref{why}),
\[
Y_{jl}^{(BA)}=\rho_{1}\delta_{jl}+\rho_{2}\hat{d}_{j}^{(BA)}\hat{d}_{l}%
^{(BA)},
\]
where $\rho_{1}=2.\,\allowbreak31$ and
$\rho_{2}=-1.\,\allowbreak58.$

We can also evaluate all of the coefficients involved in the
tensor formulas introduced earlier. So we have $\alpha_{1}=1.94,$
$\alpha_{2}=0.24,$ $\omega_{1}=0.78,$ $\omega_{2}=-0.16,$
$\psi=2.\,\allowbreak02,$ $\widetilde{\alpha}_{2}=-0.76,$ and
$\widetilde{\omega}_{1}=-0.22.$

Using equations (\ref{Fifth_part}), (\ref{ARE}), and (\ref{INTG}),
we derive
\begin{equation}
S_{jmn}^{(BA)}=a_{1}\hat{d}_{m}^{(BA)}\hat{d}_{j}^{(BA)}\hat{d}_{n}%
^{(BA)}+a_{2}\left(  \delta_{mj}\hat{d}_{n}^{(BA)}+\delta_{jn}\hat{d}%
_{m}^{(BA)}\right)  +a_{3}\delta_{mn}\hat{d}_{j}^{(BA)} \label{Ess}%
\end{equation}
where $a_{1}=-0.32,$ $a_{2}=-0.33,$ and $a_{3}=-0.0\,\allowbreak3$
and
\begin{equation}
H_{jl}^{(BA)}=b_{1}\delta_{jl}+b_{2}\hat{d}_{j}^{(BA)}\hat{d}_{l}^{(BA)},
\label{aitch}%
\end{equation}
where $b_{1}=4.\,\allowbreak25$ and $b_{2}=-2.\,\allowbreak34.$
Then, from equation (\ref{AJFluct}),
\begin{align*}
&  \overline{A_{ji}^{(BA)\prime}J_{imn}^{(BA)\prime}}\\
&  =K_{N}^{2}\left[  \kappa_{1}\hat{d}_{m}^{(BA)}\hat{d}_{j}^{(BA)}\hat{d}%
_{n}^{(BA)}+\kappa_{2}\left(
\delta_{jm}\hat{d}_{n}^{(BA)}+\delta_{nj}\hat
{d}_{m}^{(BA)}\right)
+\kappa_{3}\delta_{nm}\hat{d}_{j}^{(BA)}\right]  ,
\end{align*}
where $\kappa_{1}=-0.31,$ $\kappa_{2}=-0.02$, and
$\kappa_{3}=0.15$. This permits the calculation of the second term
in equation (\ref{final_fluc_sol}).
In a similar way, from equation (\ref{AAFluct}),%
\[
\overline{A_{js}^{(BA)\prime}A_{sl}^{(BA)\prime}}=K_{N}^{2}\left(
\eta
_{1}\delta_{jl}+\eta_{2}\hat{d}_{j}^{(BA)}\hat{d}_{l}^{(BA)}\right)
,
\]
where $\eta_{1}=0.49$ and $\eta_{2}=0.0\,\allowbreak3.$ Then, with
this and equation (\ref{JBAR}), the last term in equation
(\ref{final_fluc_sol}) is
\begin{align*}
&  \left(  \overline{A_{jk}^{(BA)}}\right)  ^{-1}\left(  \overline
{A_{sp}^{(BA)}}\right)  ^{-1}\left(
\overline{A_{li}^{(BA)}}\right)
^{-1}\overline{A_{ks}^{(BA)\prime}A_{pl}^{(BA)\prime}}\ \overline
{J_{imn}^{(BA)}}\\
&  =\psi^{-3}\left[  \xi_{1}\hat{d}_{j}^{(BA)}\hat{d}_{n}^{(BA)}\hat{d}%
_{m}^{(BA)}+\xi_{2}\left(
\delta_{jm}\hat{d}_{n}^{(BA)}+\delta_{jn}\hat
{d}_{m}^{(BA)}\right)
+\xi_{3}\delta_{nm}\hat{d}_{j}^{(BA)}\right]  ,
\end{align*}
where $\xi_{1}=0.40,$ $\xi_{2}=-0.08,$ and
$\xi_{3}=-0.0\,\allowbreak8.$ Finally, the first term in equation
(\ref{final_fluc_sol}) is the simple
average%
\begin{align*}
\overline{\left(  A_{ji}^{(BA)}\right)
^{-1}}\overline{J_{imn}^{(BA)}}  &
=\psi^{-1}\left[  \omega_{1}\hat{d}_{j}^{(BA)}\hat{d}_{m}^{(BA)}\hat{d}%
_{n}^{(BA)}\right. \\
&  \left.  +\omega_{2}\left(
\delta_{jn}\hat{d}_{m}^{(BA)}+\delta_{jm}\hat
{d}_{n}^{(BA)}+\delta_{nm}\hat{d}_{j}^{(BA)}\right)  \right]
\end{align*}
that follows from equations (\ref{AINV}) and (\ref{JBAR}).

\section{References}

Cundall, P.A., 1988. Computer simulations of dense sphere
assemblies. In: Satake, M., Jenkins, J.T. (Eds.), Micromechanics
of Granular Materials. Elsevier, Amsterdam, pp. 113-123.

Digby, P.J., 1981. The effective elastic moduli of porous granular
rocks. Journal of Applied Mechanics 48, 803-808.

Jenkins, J.T., 1997. Inelastic behavior of random arrays of
identical spheres. In: Fleck, N.A. (Ed.), Mechanics of Granular
and Porous Materials. Kluwer, Amsterdam, pp. 11-22.

Jenkins, J.T., Cundall, P. A., Ishibashi, I., 1989.
Micromechanical modeling of granular materials with the assistance
of experiments and numerical simulations. In: Biarez, J., Gourves,
R. (Eds.), Powders and Grains. Balkema, Rotterdam, pp. 257-264.

Love, A.E.H., 1927. A Treatise on the Mathematical Theory of
Elasticity. Cambridge University Press, Cambridge.

Makse, H.A., Gland, N., Johnson, D.L., Schwartz, L.M., 1999. Why
effective medium theory fails in granular materials. Physical
Review Letters 83, 5070-5075.

Norris, A.N., Johnson, D.L., 1997. Nonlinear elasticity of
granular media, Journal of Applied Mechanics 64, 39-49.

Paine, A.C., 1997. Calculation of the effective moduli of a random
packing of spheres using a perturbation of the uniform strain
approximation. In: Behringer, R.P. and Jenkins, J.T. (Eds.)
Powders and Grains 97. Balkema, Amsterdam, pp. 291-294.

Trentadue, F., 2001. A micromechanical model for a non-linear
elastic granular material based on local equilibrium.
International Journal of Solids and Structures 38, 7319-7342.

Walton, K., 1987. The effective elastic moduli of a random packing
of spheres. Journal of the Mechanics and Physics of Solids 35,
213-226.

\end{document}